\documentclass[twocolumn,preprintnumbers,amsmath,amssymb,prb,superscriptaddress]{revtex4-1}

\usepackage{graphicx}
\usepackage{float}
\usepackage{dcolumn}
\usepackage[tight]{subfigure}
\usepackage{amsmath}
\usepackage{verbatim}
\usepackage{units}
\usepackage[usenames,dvipsnames]{color}
\usepackage{bm}
\usepackage[normalem]{ulem}
\usepackage{enumerate} 
\usepackage{gensymb}

\usepackage{t1enc}
\usepackage[english]{babel}
\addto\captionsenglish{}

\usepackage[normalem]{ulem}

\newcommand{\todo}[1]{ \color{red}$\bm{\Rightarrow}$  #1 \color{black}}

\renewcommand{\todo}[1]{}

\usepackage{epstopdf}

\begin{document}

\title{Observing Magnetic Anisotropy in Electronic Transport through Individual Single-Molecule Magnets}

\author{E. ~Burzur\'{\i}}\email{E.BurzuriLinares@tudelft.nl}
\author{R. ~Gaudenzi}
\author{H. S. J. ~van der Zant}
\affiliation{Kavli Institute of Nanoscience, Delft University of Technology, PO Box 5046, 2600 GA Delft, The Netherlands}

\begin{abstract}

We review different electron transport methods to probe the magnetic properties, such as the magnetic anisotropy, of an individual Fe$_{4}$ SMM. The different approaches comprise first and higher order transport through the molecule. Gate spectroscopy, focusing on the charge degeneracy-point, is presented as a robust technique to quantify the longitudinal magnetic anisotropy of the SMM in different redox states. We provide statistics showing the robustness and reproducibility of the different methods. In addition, conductance measurements typically show high-energy excited states well beyond the ground spin multiplet of SMM. Some of these excitations have their origin in excited spin multiplets, others in vibrational modes of the molecule. The interplay between vibrations, charge and spin may yield a new approach for spin control.

\end{abstract}

\maketitle

\section{\label{sec:sect1}Introduction}

Molecular electronics aims at using individual molecules as building blocks of electronic circuits, such as molecule-scale current rectifiers, diodes or memory elements. Thanks to the versatility of synthetic chemistry, custom-made molecules can be made with a wide variety of built-in properties. Magnetic molecules can combine the molecular spin degree of freedom and its interaction with the charge carriers to process and store information, giving birth to the field of molecular spintronics \cite{Bogani2008,Sanvito2011}. This complementary path to standard spintronics, more focused on using the spin of the charge carriers, proposes to improve the spin coherence length and time thanks to the reduced size of the molecules. Single molecule magnets (SMMs), with high spin and magnetic anisotropy, are among the most promising magnetic molecules for molecular spintronics. Memory elements and spin filters are some possible applications but also the quantum nature of the spin may allow new functionalities in the field of quantum computation\cite{Loss2001,Thiele2014}. From a fundamental research point of view, SMM in electronic devices are very interesting candidates to study the interplay between charge and spin at the single molecule level.

In spite of the extensive knowledge acquired on crystals of SMM, there is still limited information of how these individual molecules preserve their magnetic properties when attached to metallic components (surfaces, electrodes...) and how these magnetic properties can be read through their interaction with the electrons. It has been reported that the Fe$_{4}$ SMM preserves its magnetic properties when grafted onto gold\cite{Mannini2009,Perfetti2014}. Moreover, these magnetic properties can be quantified and reversibly modified by changing the charge state of the molecule\cite{Zyazin2010,Zyazin2011,Burzuri2012}. Recently, the read-out of the nuclear spin flip\cite{Vincent2012} and Rabi oscillations\cite{Thiele2014} of a TbPc$_{2}$ double decker molecule have been reported.

Here, we review the main methods used to measure and quantify the magnetic properties of SMMs, mainly anisotropy and changes in spin, through their interaction with an electrical current in a three-terminal solid state device. The manuscript is divided as follows. In Section~\ref{sec:sect2} we describe the magnetic properties of SMM with special focuss on the Fe$_{4}$ SMM. In Section~\ref{sec:sect3} we summarize the different methods to measure transport through magnetic molecules and basic concepts of molecular electronics. In Sections~\ref{sec:sect4} and ~\ref{sec:sect5} we show the detection of the zero-field splitting of a SMM in the sequential electron transport and the co-tunneling regime respectively. In Sections ~\ref{sec:sect6} and ~\ref{sec:sect7}, we describe the gate spectroscopy technique for axial and transverse magnetic anisotropy. Section ~\ref{sec:sect8} treats the Kondo physics in a SMM. Finally, in Sections ~\ref{sec:sect9} and ~\ref{sec:sect10} we discuss excitations that originate from excited spin multiplets and vibrations.

\section{\label{sec:sect2}Magnetism of the Single-Molecule Magnets.}

\begin{figure}
 \includegraphics[width=0.5\textwidth]{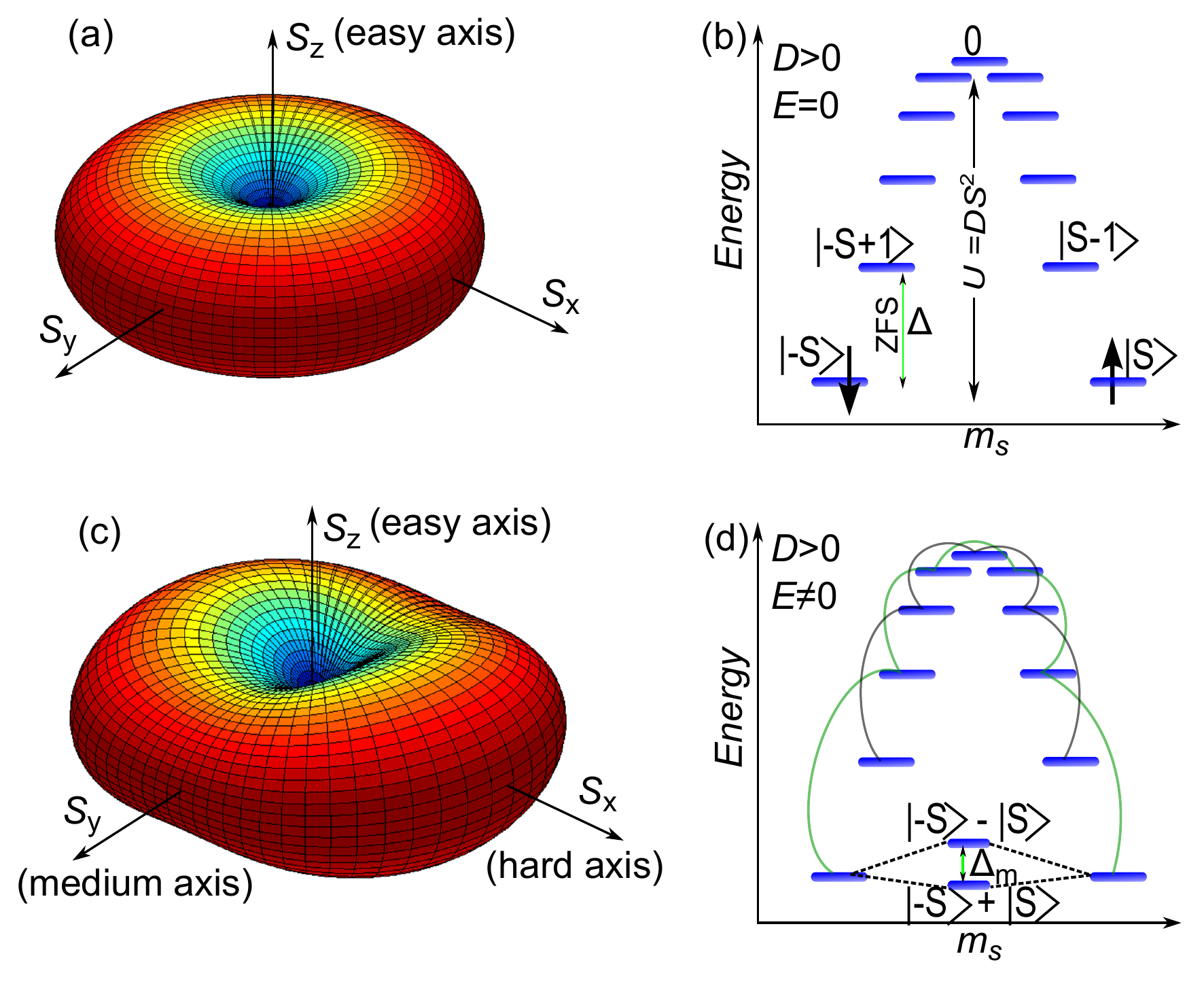}
 \caption{(a) Potential energy surface including only axial anisotropy at $B=0$. The radial distance to the surface represents the energy of a spin as a function of the orientation. An \emph{easy} magnetization axis (minimum potential) is defined along the $z$ direction. (b) Spin ground multiplet of an integer spin SMM including only axial anisotropy. (c) Potential energy surface including the transverse anisotropy. A \emph{hard} and a \emph{medium} axes for spin reversal are defined. (d) Spin ground multiplet of an integer spin SMM including axial and transverse magnetic anisotropy. The solid lines connect the states that are mixed by the transverse anisotropy.}
 \label{figure1}
 \end{figure}

Transition-metal based single-molecule magnets are generally made of several magnetic ions coupled together by exchange $J$ or superexchange interactions. At low temperatures ($k_{\textrm{B}}T\lesssim J$) the magnetic behaviour can be described by an effective total spin $S$ that results from the vectorial addition of the spins ($s$) of the individual magnetic ions. This is known as the giant-spin approximation. The molecular field defined by the symmetry of the molecule gives rise to a magnetic anisotropy that creates an \emph{easy magnetization axis} as sketched in Fig.\ref{figure1}(a). The degeneracy of the spin ground multiplet breaks into $2S+1$ states ($m_s=-S,-S+1,...,0,...,S-1,S$) as depicted in Fig.\ref{figure1}(b). The molecule can be modeled by a so-called "giant-spin Hamiltonian":

\begin{equation}
H=-DS_{z}^{2}+ E(S^{2}_{x}-S^{2}_{y})-g\mu_\textrm{B}\overrightarrow{S}\cdot\overrightarrow{B}
\label{GS}
\end{equation}

\noindent where the first term describes the axial anisotropy of the molecule. The second term characterizes the transverse anisotropy as second order perturbations to the axial anisotropy. The last term describes the Zeeman interaction of $S$ with an external magnetic field $B$, where $g$ is the Land\'{e} factor and $\mu_{\textrm{B}}$ is the Bohr magneton. $S_\text{z}$ is the projection of the spin along the easy axis 'z' and $D$ is the axial anisotropy parameter. The angle between $B$ and the easy axis 'z' is defined as $\theta$. The energy levels $m_\text{s}$ are distributed over a barrier $U=DS^{2}$ that separates the "up" and "down" orientations of the spin (see Fig.\ref{figure1}(b)). $S_x$ and $S_y$ are the spin projections on a \emph{hard} and \emph{medium} axes respectively (see Fig.\ref{figure1}(c)) and $E$ is the transverse anisotropy parameter. It mixes the $m_s$ states of both sides of the barrier and breaks the degeneracy of the $\pm m_{s}$ doublets by $\Delta_m$ leading to a small tunnel probability between $m_s$ and $-m_s$ states if $S$ is integer as sketched in Fig.\ref{figure1}(d).

\begin{figure}
 \includegraphics[width=0.5\textwidth]{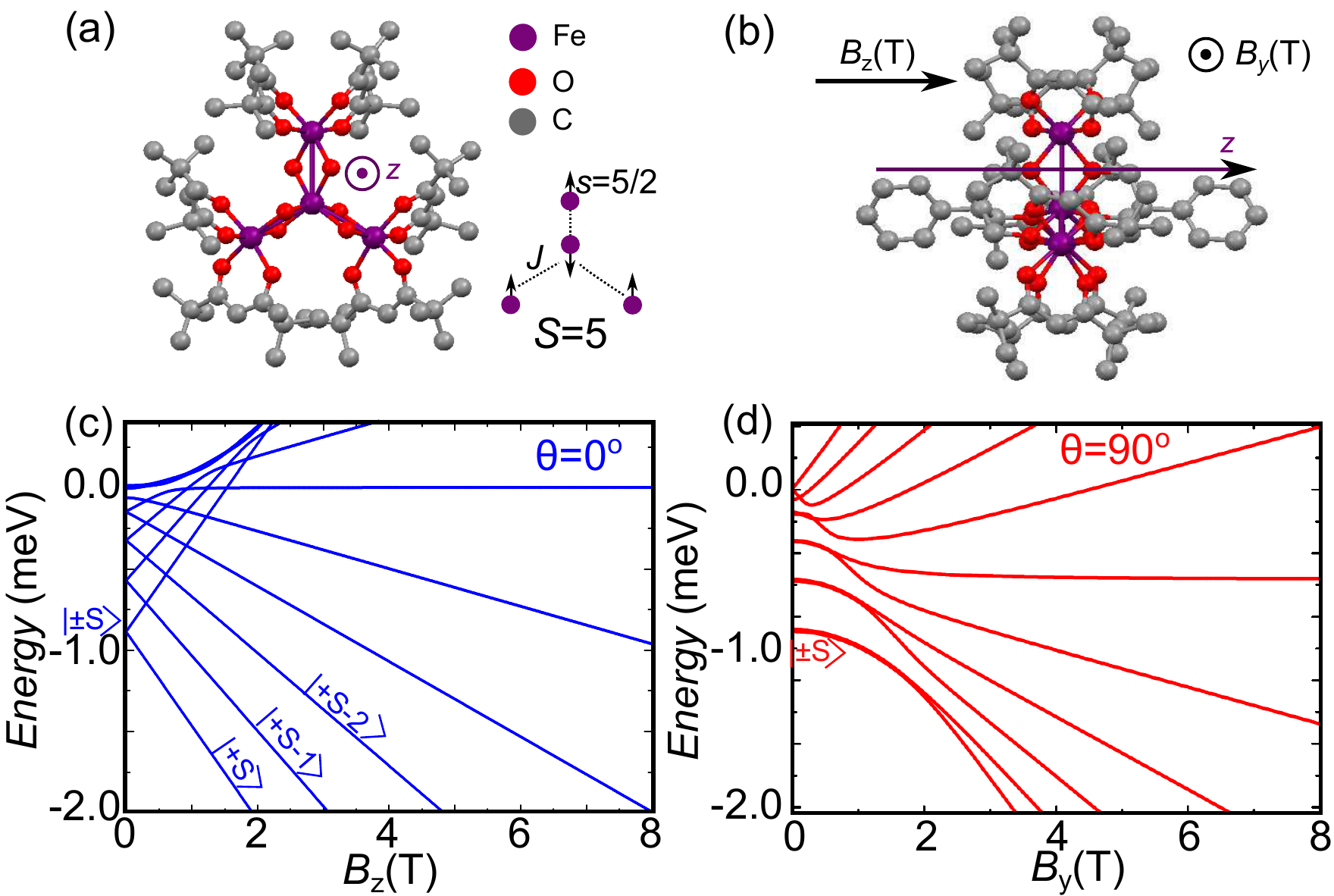}
 \caption{(a) Schematics of the magnetic core of the Fe$_4$ SMM. Three peripheral Fe$^{3+}$ ions are antiferromagnetically coupled to a central Fe$^{3+}$ ion giving a total spin $S_N=5$ in the neutral charge state $N$. Phenyl rings, peripheral ligands and hydrogen atoms are removed for clarity. (b) Lateral view of the Fe$_{4}$ SMM showing the phenyl rings that, ideally, contact with the gold electrodes.~\textit{Bottom}: Energy of the spin ground multiplet of the Fe$_{4}$ SMM when $B$ is parallel (c) and  perpendicular (d) to the easy axis. The parameter $\theta$ defines the angle between $B$ and the easy axis 'z'.}
 \label{figure2}
 \end{figure}

In this review, we focus on the Fe$_{4}$ SMM with formula [Fe$_{4}$(L)$_{2}$(dpm)$_{6}]\cdot$ Et$_{2}$O where Hdpm is 2,2,6,6-tetramethyl-heptan-3,5-dione and H3L is the tripodal ligand 2-hydroxymethyl-2-phenylpropane-1,3-diol, which carries a phenyl ring\cite{Accorsi2006}. The magnetic core is made of 3 peripheral Fe$^{3+}$ ions ($s=5/2$) antiferromagnetically coupled to the central Fe$^{3+}$ as schematically shown in Fig.\ref{figure2}(a). The resulting spin ground state in the neutral charge state $N$ is $S_N=5$. The easy axis of the molecule points along the phenyl rings direction, that is, perpendicular to the plane containing the four Fe$^{3+}$ ions as shown in Fig.\ref{figure2}(b). The first excited spin multiplet ($S=4$) lies around 5 meV above the spin ground multiplet and therefore the giant spin Hamiltonian is a good approximation at our experimental temperatures, typically below 1.8 K (0.15 meV). The Fe$_{4}$ SMMs are not functionalized with any specific group to link to the Au electrodes (like thiol or thiophene). The molecule-electrode interaction is therefore mediated only by weak van der Waals interactions ideally through the phenyl rings.

Magnetic properties of Fe$_{4}$ SMM crystals have been extensively studied with different spectroscopic techniques such as Electron Paramagnetic Resonance EPR, neutron scattering and superconducting quantum interference device SQUID magnetometry. From these measurements the bulk values of the anisotropy parameters have been determined: $D=56~\mu$eV and $E=2.85~\mu$eV yielding a ZFS of $0.5$ meV and an energy barrier $U=1.4$ meV~\cite{Accorsi2006}. These techniques, however, are only sensitive to large ensembles of molecules and are not valid for the individual SMM. As a result of this limitation, little is known about the magnetic anisotropy of individual and charged SMMs. The Fe$_{4}$ SMM, however, has proven to preserve its magnetic properties in solution \cite{Schlegel2010} when attached to gold surfaces \cite{Mannini2009, Malavolti2014}, electrodes\cite{Zyazin2010,Zyazin2011} or nanoparticles\cite{Perfetti2014}.

\section{\label{sec:sect3} Electronic transport through individual magnetic molecules.}

\begin{figure}
 \includegraphics[width=0.5\textwidth]{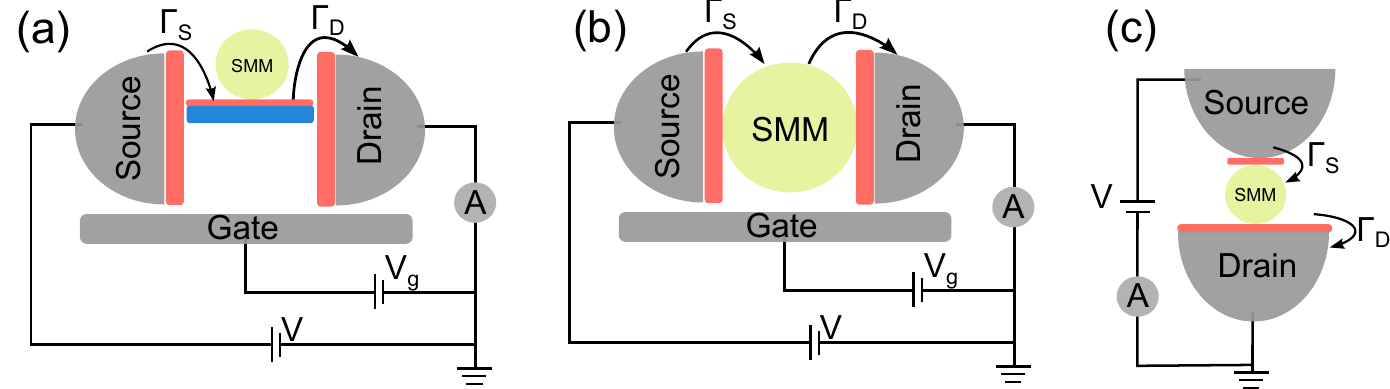}
 \caption{Different approaches to electronically probe the magnetic properties of an SMM. (a) \textbf{Indirect approach:} The charge carriers flow through an intermediate path that is coupled to the molecule. (b) \textbf{Direct approach: spin transistor.} The charge carriers flow through the magnetic core of the molecule. This approach allows to change the redox state of the molecule with a gate voltage more easily. (c) \textbf{Direct approach: STM configuration}. The tip of an STM is used as source and a metallic substrate is used as drain electrode. It allows a local probing of the molecule.}
 \label{figure4}
 \end{figure}

We distinguish two different approaches to measure electronic transport through a single molecule depending on the strength of the coupling between the charge carriers and the spins of the SMM. In the \emph{indirect} approach, the charge carriers do not flow through the magnetic core of the molecule. Instead, electrons flow through a close-by intermediate transport channel\cite{Urdampilleta2011,Candini2011,Vincent2012} to which the molecule is linked as shown in Fig.\ref{figure4}(a). This has been achieved using carbon nanotubes or big organic shells surrounding the magnetic core that are in turn attached to the source/drain electrodes via tunnel barriers. It is the less invasive of the two approaches and it does not allow to change the redox state of the molecule easily. In contrast, in the \emph{direct} approach described in this paper, the current flows through the magnetic core of the molecule that is directly coupled to the source and drain electrodes via tunnel barriers (see Fig.\ref{figure4}(b)).

The Scanning Tunnel Microscope STM configuration depicted in Fig.\ref{figure4}(c) is an example of the direct approach. The tip of the STM is used as the source electrode whereas a metallic substrate acts as drain electrode. The advantages of the STM approach are the in-situ imaging of the molecule\cite{Heinrich2013}, the local electrical probing on different parts of the molecule \cite{Miyamachi2012} and a higher control over the electronic coupling $\Gamma$. On the other hand, a limitation is the absence of a gate electrode. Moreover most molecules hybridize with the surface thereby altering their complex magnetic structure as observed in recent examples\cite{Mannini2008}. This can be solved by using more robust SMM like the Fe$_4$.

\begin{figure}
\includegraphics[width=0.5\textwidth]{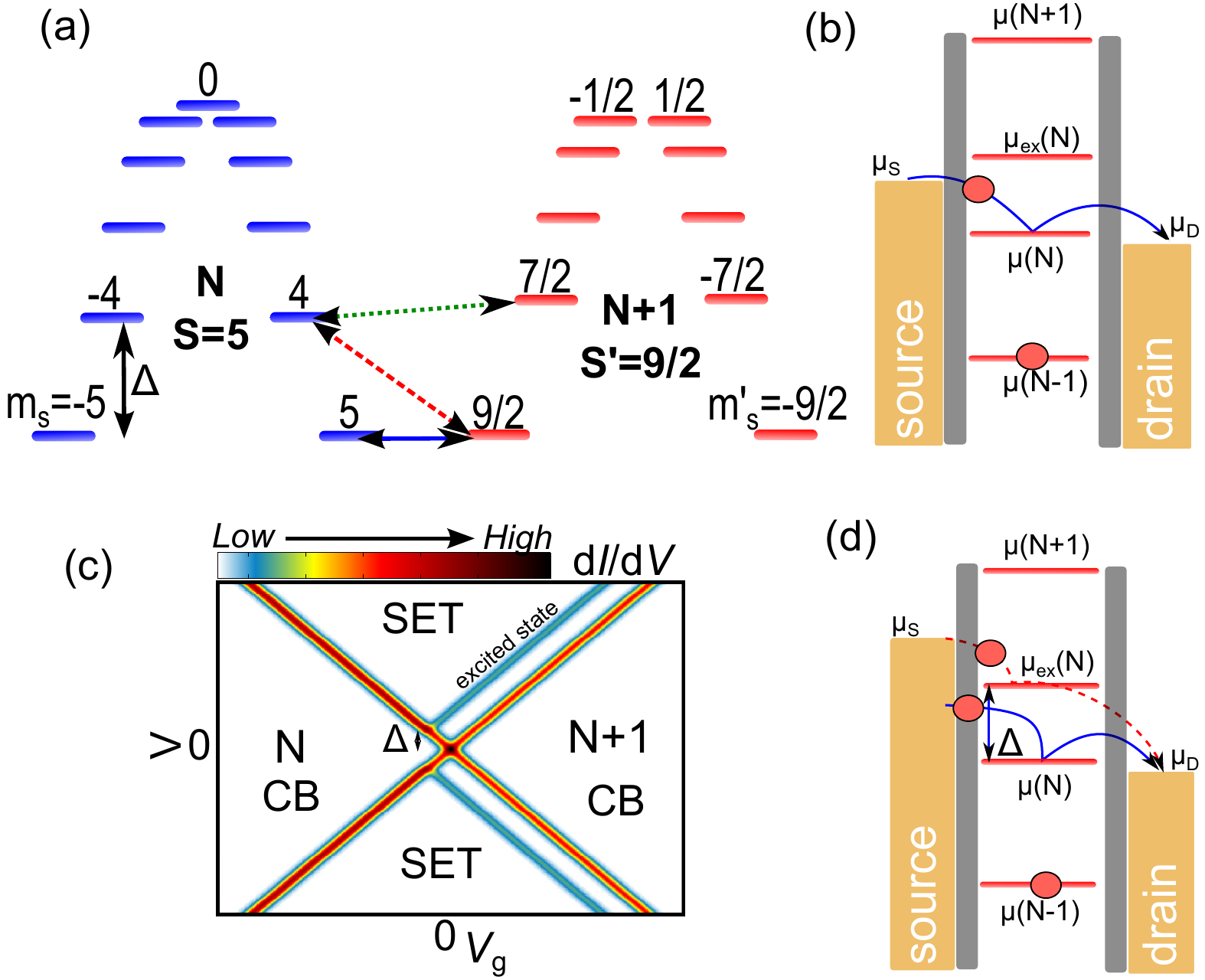}
\caption{(a) Energy scheme displaying the ground spin multiplet for the Fe$_{4}$ SMM in the neutral $N$ and the reduced $N+1$ charge states. We only represent the $N+1$ state corresponding to $S=9/2$ for clarity. Arrows indicate ground to ground (blue), ground to excited (red) and excited to excited states (green) transitions. (b) Electrochemical scheme of a first-order sequential tunneling process involving ground states. (c) Schematic $\text{d}I/\text{d}V$ color plot as a function of $V$ and $V_\text{g}$. First order transport is included. (d) Electrochemical scheme of a first-order sequential tunneling process involving excited states.}
\label{figureSET}
\end{figure}

A molecule in the three-terminal devices used hereafter can be schematically depicted as in Fig.\ref{figure4}(b). The molecule bridges the nanogap formed between source and drain gold electrodes by self-breaking electromigration\cite{Park1999,ONeill2007} and is capacitively coupled to an aluminum gate electrode. By means of the electrostatic field generated by the gate, it is possible to reduce/oxidize the molecule, i.e., to add/substract one electron to/from the molecule. The energy needed, for instance, to bring the molecule from its charge state $N$ and total spin state $S$ to its charge state $N+1$ and total spin state $S'$, as shown in Fig.\ref{figureSET}(a) for the Fe$_{4}$, is defined by the electrochemical potential $\mu(N+1) \equiv E(N+1,S', m_s')-E(N,S, m_s)$, where $\Delta S \equiv S' - S = \pm 1/2$ and $\Delta m_s \equiv m_s' - m_s = \pm 1/2$ are set by conservation of angular momentum. The electrochemical potentials of the electrodes can be defined as $\mu_{\text{S}}=\mu_\textrm{F}-eV/2$ and $\mu_{\text{D}}=\mu_\textrm{F}+eV/2$ for source and drain respectively, where $\mu_F$ is the Fermi level and $V$ is a bias voltage. Applying $V$ and gate $V_\text{g}$ voltages, the relative alignment between the electrochemical potential of the electrodes and the molecular states $\mu (N)$, $\mu (N+1),...$  can be varied. An electrochemical scheme of the molecule and the electrodes is shown in Fig.\ref{figureSET}(b). If the condition for resonant transport is met ($\mu_F + eV/2 > \mu (N+1) + \beta eV_\textrm{g} > \mu_F - eV/2$) electrons can hop one at a time from the source to the molecule and from the molecule to the drain, in a two-step process. Transport in this regime is denominated single-electron tunneling (SET) transport and takes place according to the cycle $(N, S, m_\text{s}) \rightarrow (N+1, S', m_s') \rightarrow (N,S,m_\text{s})$. Upon reaching the resonant transport condition, the differential conductance ($\text{d}I/\text{d}V$) shows a peak and the current through the molecule has a step-like increase.

A typical representation of the result of a measurement is shown in Fig.\ref{figureSET}(c); it is a color plot of the $\text{d}I/\text{d}V$ as a function of $V$ and $V_{\text{g}}$. The red slanted lines, known as Coulomb edges, are due to transitions from the ground state of one charge state to the one of an adjacent charge state. The crossing point at zero bias is often called degeneracy point or Coulomb peak. In the case of the Fe$_{4}$ SMM, the transition may correspond to the ($S=5$, $m_\text{s}=\pm5$) and  ($S'=9/2$, $m'_\text{s}=\pm9/2$) or ($S'=11/2$, $m'_\text{s}=\pm11/2$) marked with a blue arrow in Fig.\ref{figureSET}(a). First-order tunneling transport is energetically allowed in the high-bias regions of the conductance map confined by those edges, indicated by SET.

We have measured around 300 junctions at low temperatures ($T<1.8$ K). Most of them do not show any gate dependence or are very high ohmic junctions ($>1~\textrm{G}\Omega$) indicating that the junctions are empty. Interestingly, 31 junctions exhibited gate dependence or steps in $I-V$ measurements. From these, 16 displayed a clear Coulomb peak separating two stable charge states\cite{footnote1}. The state-of-the-art gating technology with solid gates is limited by the maximum voltage supported by the gate and the dielectric and the efficiency of the coupling $\beta$ between the electrical field and the molecule. The value of $\beta$ is experimentally obtained as $\beta=\left(|\alpha_{+}|^{-1}+|\alpha_{-}|^{-1}\right)^{-1}$ where  $\alpha_{+}$ and $\alpha_{-}$ are respectively the positive and negative slopes of the Coulomb edges. Typical values are $\beta=0.03-0.10$ which allow to shift the energy levels by hundreds of meV. Given the fact that charging energies of small molecules are expected to be of the order of eV, $\text{d}I/\text{d}V$ color maps therefore are expected to show two stable charge states separated by an SET region. The position of the degeneracy point is not a priori determined as it shifts in gate voltage from sample to sample what makes difficult to assign the neutral charge state to a specific Coulomb diamond. The different magnetic behavior of the different charge states can be still used, in some cases, to assign the neutral charge state as explained in the next sections.

The electronic coupling of the molecule to the source $\Gamma_\textrm{S}$ and drain $\Gamma_\textrm{D}$ electrodes defines the rate at which the electrons tunnel through the molecule. $\Gamma$ can be then defined as $\Gamma=\Gamma_\textrm{D}+\Gamma_\textrm{S}$. At very low temperatures ($k_\textrm{B}T\ll \Gamma$) this electron coupling is the main contribution to the broadening of the Coulomb edges and peak. The value of $\Gamma$ can therefore be estimated from the FWHM of the Coulomb peak at zero bias according to $\Gamma=\textrm{FWHM}\cdot\beta$. This value depends on the specific arrangement of the molecule between the electrodes and can not be controlled in solid-state devices where the electrodes are fixed. In the case of the Fe$_4$ SMM, we found that $\Gamma$ ranges from 0.5 meV to 3.8 meV in the different samples. It is worth noting that we observe asymmetric coupling ($\Gamma_\textrm{S}/\Gamma_\textrm{D}\neq1$) in $84\%$ of the measured Fe$_4$ SMM junctions.

\section{\label{sec:sect4}Zero-field splitting in the SET regime.}

Excited states can contribute to electron transfer once their chemical potential ($\mu_\textrm{ex} (N)$) lies within the bias window, see red arrows in Fig.\ref{figureSET}(d). Depending on the initial and final states, they are referred to as ground-to-excited, excited-to-ground (red arrows in Fig.\ref{figureSET}(a)) or excited-to-excited (green arrow in Fig.\ref{figureSET}(a)) state transitions and follow cycles of the kind $(N, S, m_\text{s,i}) \rightarrow (N+1, S', m'_{s}) \rightarrow (N,S,m_\text{s,f})$, where $i$ and $f$ denote initial and final $m_{s}$ states that fulfill $m_\text{s,f} - m_\text{s,i} =0, \pm1$ and $m'_{s}$ is a state that fulfills $m'_{s}-m_{s,i}=\pm1/2$. These transitions manifest themselves as additional lines running parallel to the transport edges within the SET regions that intersect the opposite Coulomb edges at finite bias (see Fig.\ref{figureSET}(c)). The intersecting bias gives the energy of the excitation $\Delta$ with respect to the ground state. Moreover, from the direction and position of these transport lines, it is possible to determine to which charge state the excitations belong. For instance, the transport lines terminating in the $N$ charge-state Coulomb diamond indicated in Fig.\ref{figureSET}(c) are related to the energy difference $\Delta = E(N,S,m_{\text{s,i}})-E(N,S,m_{\text{s,f}})$ between two $m_\text{s}$-spin states belonging to the spin $S$-multiplet of the $N$ charge state.

 \begin{figure}
 \includegraphics[width=.5\textwidth]{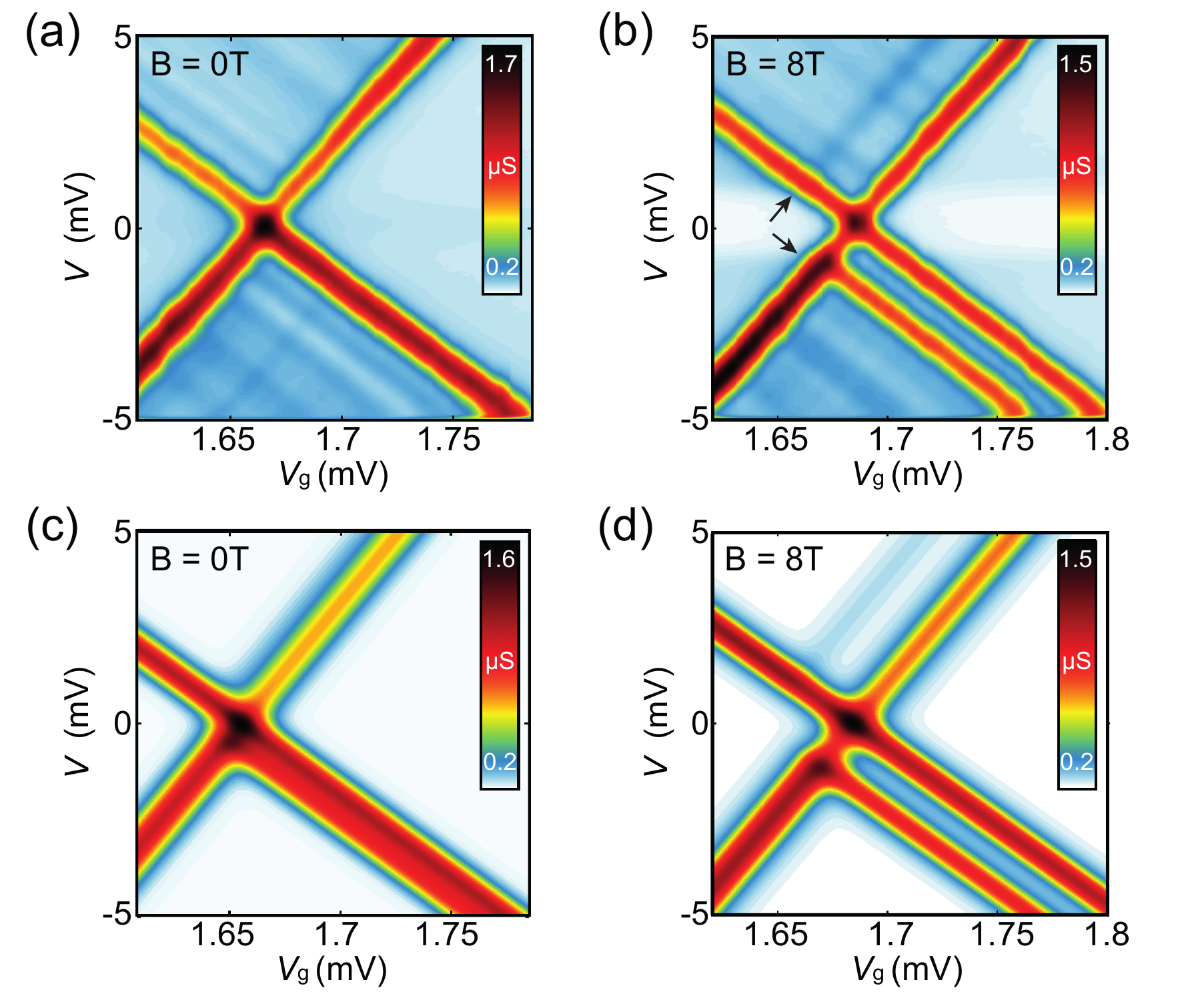}
 \caption{ (a) $\textrm{d}I/\textrm{d}V$ colour plot of an individual Fe$_4$ SMM junction as a function of $V$ and $V_\textrm{g}$ at $B = 0$ T and $T=360$ mK (Sample AB). Coulomb-blockade is visible on either side of the degeneracy peak, marking two regions with a definite charge state. Faint inelastic co-tunneling that can be attributed to the ZFS lines are also visible. Within the SET transport regions, excitations of vibronic origin appear. The asymmetry in the couplings to the electrodes is evident in the different intensities of the Coulomb edges. (b) Same colour plot as in (a) measured at $B = 8$ T and $T=360$ mK. The two Coulomb edges at positive and negative biases (black arrows) together with the corresponding ZFS inelastic cotunneling excitations are now split. (c) $\textrm{d}I/\textrm{d}V$ colour plot calculated with a three-state master equation model at $B = 0$ T. (d) Same as in (c) calculated at $B = 8$ T. In both simulations, the different conductance intensities of the Coulomb edges can be reproduced with a ratio $\Gamma_\textrm{S}/\Gamma_\textrm{D} = \Gamma '_\textrm{S}/\Gamma '_\textrm{D} \approx 4$, where the prime indicates excited state related quantities.}
 \label{figure5}
 \end{figure}

In Fig.\ref{figure5}(a) the $\textrm{d}I/\textrm{d}V$ map of an Fe$_{4}$ junction (sample AB) as a function of $V$ and $V_{\textrm{g}}$ measured at $T=360$ mK and a magnetic field $B=0$ T is shown. Two low-conductance regions on either side of the degeneracy point at gate voltage $V_{\textrm{g}} \approx  1.67$ V can be distinguished. These two Coulomb-blockaded regions can be associated with the two stable states ($N = 0, S = 5$) and ($N+1 = 1, S' = 9/2$) for the lower and higher gate voltages respectively (determination of the states will be explained in Sect.\ref{sec:sect6}). Figure \ref{figure5}(b) shows the conductance map at the same temperature and at an applied magnetic field $B=8$ T. The edges marking the transition between the SET transport and the $N+1$ charge state have split in two pairs of lines (indicated by arrows). The excited-to-ground state excitation terminates now at $V \approx \pm 1.1 $ meV. The increase of the excitation energy is consistent with the evolution of the zero-field splitting in a magnetic field. Assuming $\Delta(B = 0 \textrm{T}) = 0.5$ meV as in bulk, the splitting of the lines and the energy $\Delta(B = 8 \textrm{T}) = 1.1$ meV can be reproduced for $\theta \approx 50 \degree$ in equation (\ref{GS}). This angle is confirmed by gate spectroscopy measurements performed on the sample AB that are described in Sect.\ref{sec:sect6}. Note that because of the lifetime broadening $\Gamma \approx 0.68 \textrm{ meV} \gtrsim \Delta(B = 0 \textrm{T})$ the ZFS cannot be resolved at low fields. The analysis of the energies and the allowed spin transitions thus let us conclude that this transition follows the cycle $(N,5,\pm 5) \rightarrow (N+1,9/2,\pm 9/2) \rightarrow (N,5,\pm 4)$. \\
\indent
In order to confirm that the observed excitation corresponds to the zero-field splitting transition, differential conductance maps at various magnetic fields are simulated with a minimal three-state rate equation model. The magnetic energy levels are derived by diagonalizing the Hamiltonian in equation (\ref{GS}) for the two charged states $N$ and $N+1$ and for various magnetic fields. Transition rates, occupation probabilities of the magnetic levels and the resulting current are then calculated within the standard (stationary) master equation framework, assuming thus weak coupling between the molecule and electrodes as expected for a physisorbed molecule. Motivated by the low thermal energy and the weak mixing of the spin eigenstates at angles $\theta \lesssim 60 \degree$, the model assumes the magnetic number $m_s$ to be an approximately good quantum number and includes only the first two lowest spin eigenstates of each charge state. The results of the calculations for $B = 0$ T and $B = 8$ T at an angle $\theta = 50 \degree$ are shown in Fig.\ref{figure5}(c, d) respectively. As can be seen, the energy of the transition $(N,5, +5) \rightarrow (N,5, +4)$ increases in field consistently with the experimental data. In order to reproduce the specific conductance values observed in the experiments, asymmetric tunnel couplings to the electrodes for the states are employed. Good agreement is obtained for a ratio $\Gamma_\text{S}/\Gamma_\text{D} = \Gamma'_\text{S}/\Gamma'_\text{D} \approx 4$. An asymmetrical geometric arrangement of the molecule with respect to the source and drain electrodes can account for this ratio.  \\
\indent
Additional transport lines present in the SET are absent in the simulations. These excitations can be attributed to vibron-assisted processes, as they are roughly harmonically-spaced and do not change their energy in magnetic field. Further details on the interaction with vibrational states is given in Sect.\ref{sec:sect10}.

\section{\label{sec:sect5}Zero-field splitting in the co-tunneling regime.}

Magnetic excitations can also contribute to transport in the Coulomb blockade regime\cite{Zyazin2010}. In this case, sequential electron transport is suppressed and therefore the magnetic transitions can only take place via high-order processes like inelastic co-tunneling. The current level for a co-tunneling process is proportional to $\Gamma^{2}$ and therefore it is only relevant for intermediate or high values of $\Gamma$; i.e., $\Gamma\gtrsim k_\textrm{B}T,\Delta$ (See Ref.\cite{Thijssen2008} for a review).

 \begin{figure}
 \includegraphics[width=0.5\textwidth]{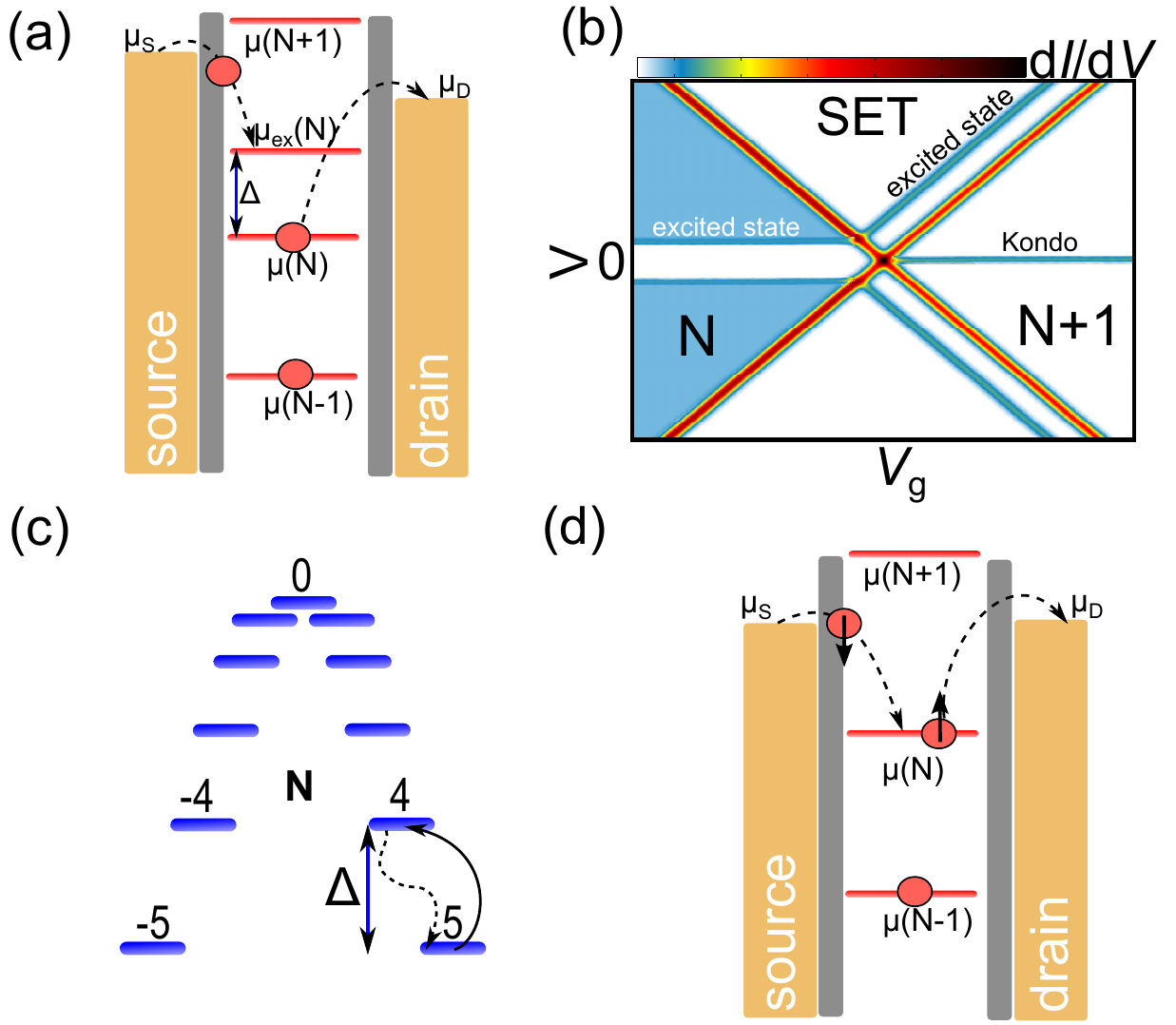}
 \caption{(a) Electrochemical scheme of an inelastic co-tunneling excitation. One electron can jump out of the molecule leaving it in a virtual "forbidden" state as long as another electron enters the molecule in a time $\Delta t < \hbar/(E_{\text{c}}+\Delta)$. (b) Stability diagram of a magnetic molecule showing different high-order tunneling processes. (c) Spin ground multiplet for the Fe$_4$ SMM showing the magnetic transition corresponding to an inelastic co-tunneling process. (d) Electrochemical scheme of a Kondo spin-flipping process.}
 \label{figure3}
 \end{figure}

Figure \ref{figure3}(a) shows the electrochemical scheme of an inelastic co-tunneling process. One electron hops from the molecule into the drain leaving the molecule in a forbidden virtual state. This event is possible if another electron jumps into the molecule in a time $\Delta t\leq\hbar/(E_\text{c}+\Delta)$ according to Heisenberg's uncertainty principle. There,  $E_\text{c}$ is the charging energy. The incoming electron can jump into a different excited state with energy $\Delta$ as long as $V\geqslant \Delta$. A step in $\text{d}I/\text{d}V$ appears at the threshold voltage $V=\Delta$ as sketched in Fig.\ref{figure3}(b) that merges with the corresponding SET excitation at the diamond edge. Two electrons are involved in this second-order process and the spin selection rules dictate that only transitions fulfilling $\Delta S=0,1$ and $\Delta m_\text{s}=0,\pm1$ are allowed.

In the Fe$_{4}$ SMM, the transition from the ground state ($S=5$, $m_\text{s}=\pm5$) to the first excited state ($S=5$, $m_\textrm{s}=\pm4$) is the most prominent transition in the co-tunnel regime, as sketched in Fig.\ref{figure3}(c); i.e., it concerns the ZFS in a single charge state where $\Delta S=0$ and $\Delta m_{s}=1$. The ZFS splitting is the largest energy difference in the spin ground multiplet $S=5$. Accordingly, once the transition is activated, all other transitions are allowed as well and therefore only one step in the $\text{d}I/\text{d}V$ is expected. At higher bias voltages, transitions to excited spin multiplets fulfilling  $\Delta S=1$ ($S=5$ to $S=4$ in the neutral state) may also become visible as discussed in Sect.\ref{sec:sect9}.

\begin{figure}
\includegraphics[width=0.5\textwidth]{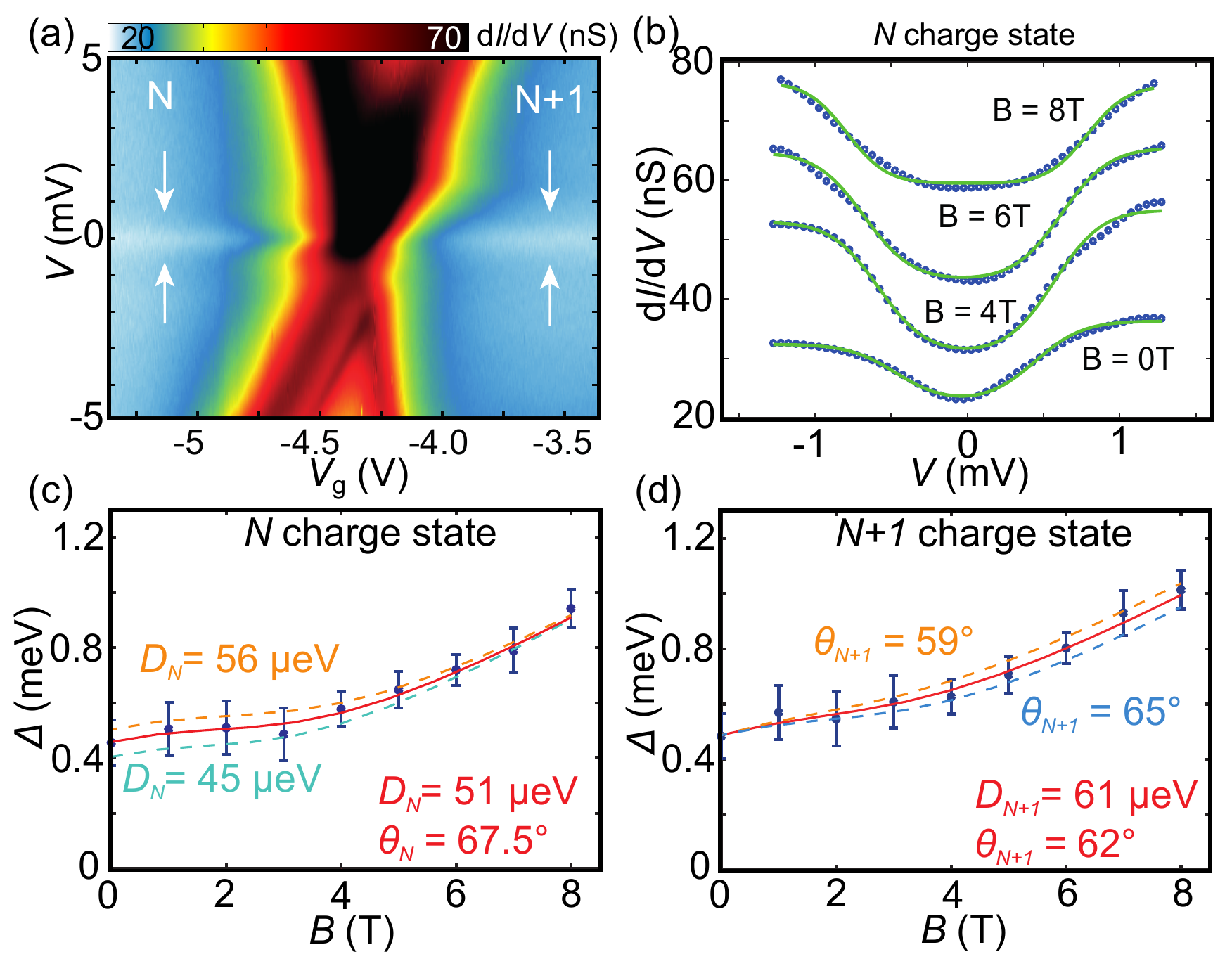}
\caption{(a) $\textrm{d}I/\textrm{d}V$ color plot as a function of $V$ and $V_{\textrm{g}}$ for an Fe$_{4}$ SMM junction (sample AH) measured at $T=570$ mK. Co-tunneling excitations appear at 0.5 mV in the left and right-hand charge states. (b) $\textrm{d}I/\textrm{d}V$ traces measured at $V_{\textrm{g}}=-5.1$ V and different vales of $B$. The energy of the excitation increases with increasing $B$ as expected for the ZFS. Solid lines are fits to a Lambe-Jaklevic formula to obtain the energy of the excitation $\Delta$. (c) and (d) $\Delta$ plotted as a function of $B$ for the left and the right charge states respectively. The solid lines are fits to equation (\ref{GS}) with parameters $D_N=51~\mu$eV, $S_N=5$, $S_{N+1}=9/2$ and $D_{N+1}=61~\mu$eV. Orange and blue dashed lines are upper and lower boundaries for $\theta_{N+1}$ and $D_{N}$. }
\label{figureCOT}
\end{figure}

Figure \ref{figureCOT}(a) shows the $\textrm{d}I/\textrm{d}V$ color plot measured at $T=570$ mK as a function of $V$ and $V_\textrm{g}$ of an Fe$_{4}$ junction (sample AH) showing again two adjacent charge states. Two parallel co-tunneling lines show up within the left-hand charge state with energy $\Delta=\pm 0.5$ meV that is close to the expected energy of the ZFS in bulk Fe$_{4}$. Two symmetric excitations appear as well within the right charge state with energy $\Delta=\pm 0.5$ meV. To gain a deeper insight in the magnetic nature of the excitations, the evolution of $\Delta$ in a magnetic field has been examined.

Figure~\ref{figureCOT}(b) shows different $\textrm{d}I/\textrm{d}V$ traces measured at $V_\textrm{g}= -5.1$ V and at different magnetic fields. The co-tunneling lines clearly move apart by increasing the magnetic field; i.e., the energy of the excitation increases as expected for the $\Delta$. The solid lines are fits of these curves to a Lambe-Jaklevic equation\cite{Lambe1968, Kogan2004}:

\begin{equation}
\begin{split}
\text{d}I/\text{d}V=A_{\textrm{e}}+A_{\textrm{i+}}F\left(\frac{-e(V+V_{\text{0}})+\Delta}{k_{\textrm{B}}T}\right)\\
+A_{\textrm{i-}}F\left(\frac{e(V+V_\textrm{0})+\Delta}{k_\textrm{B}T}\right),
\end{split}
\label{LJ}
\end{equation}

\noindent where the parameters $V_{\textrm{0}}$ and $A_{\textrm{e}}$, are horizontal and vertical offsets respectively, and $A_{\textrm{i+}}$ and $A_{\textrm{i-}}$ account for bias asymmetries in the $\textrm{d}I/\textrm{d}V$. The energy $\Delta$ obtained from these fits is plotted as a function of $B$ in Figs.\ref{figureCOT}(c) and (d) for the left and the right-hand charge state respectively. The evolution of $\Delta$ in a magnetic field provides information about the magnetic anisotropy of the molecule\cite{Zyazin2010} since the ZFS is very sensitive to $\theta$ and $D$ as seen in Fig.\ref{figure2}(c-d). The solid line in Fig.\ref{figureCOT}(c) is a fit to $\Delta$ calculated by numerical diagonalization of equation (\ref{GS}) with $S_N=5$ $D_N=51~\mu$eV and $S_{N+1}=9/2$, $\theta_N=67.5\degree$. The same analysis is performed for the right charge state in Fig.\ref{figureCOT}(d) and yields $D_{N+1}=61~\mu$eV and $\theta_{N+1}=62\degree$. The magnetic anisotropy therefore increases around $20\%$ by adding one electron to the molecule. The nonlinear dependence of the $\Delta$ with $B$ is proof of the magnetic anisotropy of the molecule and points to high values of $\theta$. This nonlinear dependence has also been observed in samples AB and B. The orange and blue dashed lines in Figs.\ref{figureCOT}(c) and \ref{figureCOT}(d) are upper and lower boundaries of the fit by using different values of $D_N$ and $\theta_{N+1}$ respectively. Note that the high-$B$ part of the dependence is more sensitive to variations in $\theta$ whereas the low-$B$ part is more sensitive to variations in $D$. This is a consequence of the predominance of the Zeeman effect over the axial anisotropy at high $B$ and viceversa.

We have observed low-bias co-tunneling excitations in 7 of the 31 junctions showing molecular signatures. The values of the excitations are summarized in Table.\ref{TableZFSCOT}. We generally find that the values are of the order of 0.5 meV with variations from sample to sample. Typical values for the ZFS are lower than 1 meV (0.55 meV in the case of bulk Fe$_4$). Additional low-bias excitations, such as zero-bias anomalies or vibrations, are often present in the measurements depending on the coupling of the molecule with the electrodes. The identification of the ZFS may therefore not be straightforward in those cases. Figure \ref{Kondo}(c) illustrates one of these examples. In that case, the ZFS appears as small shoulders on top of a zero-bias resonance due to Kondo interactions as explained in Sect.\ref{sec:sect8}.
\begin{table}
\begin{tabular}{||c||c|c||}
\hline
\hline
Sample & $\Delta ~(\pm0.1)$  (meV) (left)  & $\Delta ~(\pm0.1)$ (meV) (right)\\
\hline
\hline
A$^+$ & 0.6 & 0.9 \\
\hline
B$^+$ & 0.9 & 0.6 \\
\hline
G$^+$ & 0.5 & 0.5 \\
\hline
Y & 0.2 & 0.6\\
\hline
AB & 0.5 & 0.4\\
\hline
AD$^+$ & 0.7 & -\\
\hline
AH$^+$ & 0.5 & 0.5\\
\hline

\end{tabular}
\caption{Statistics of low-bias co-tunneling excitations ($\Delta < 1$ meV). The values of $\Delta$ of samples with '+' are obtained from the fits to the Lambe-Jaklevic equation at $B=0$ T. For the other two, $\Delta$ could not be obtained at low $B$, e.g. masked by a Kondo peak (sample Y in Fig.\ref{Kondo}(c)) or not visible at $B=0$ (sample AB in Fig.\ref{figure5}(a)). $\Delta (B=0)$ is then obtained by extrapolating high-$B$ values of $\Delta$ to $B=0$ T. Samples A and B were previously reported in Ref.\cite{Zyazin2010}. Sample G was reported in Ref.\cite{Burzuri2012}.}
\label{TableZFSCOT}
\end{table}

\section{\label{sec:sect6}Gate spectroscopy: Axial Magnetic Anisotropy.}

\begin{figure}
 \includegraphics[width=0.5\textwidth]{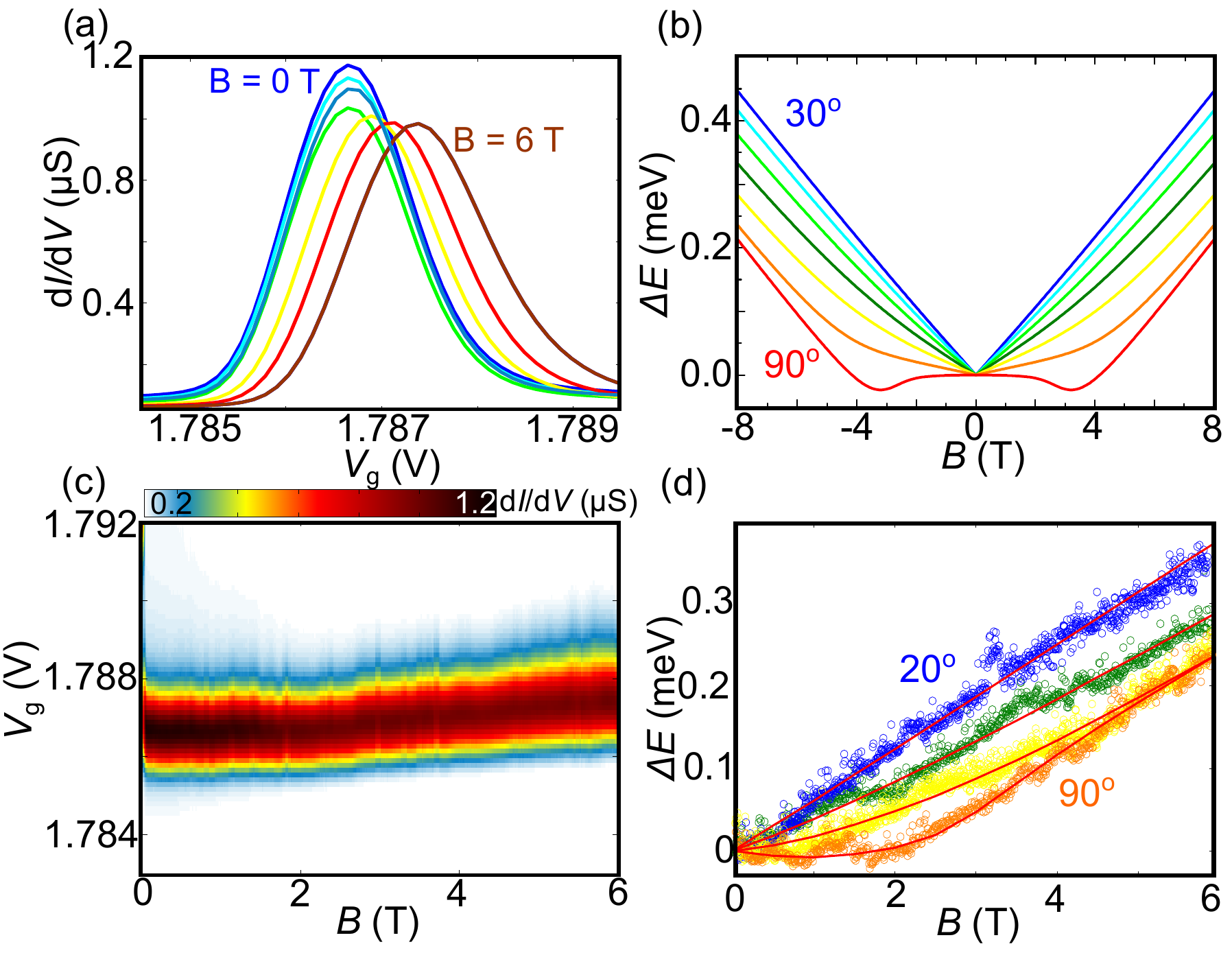}
 \caption{(a) $\text{d}I/\text{d}V$ vs $V_\textrm{g}$ at zero bias around the Coulomb peak of sample AB measured from $B=0$ T to $B=6$ T in steps of 1 T and at $T=360$ mK. The peak shifts towards higher $V_{\textrm{g}}$ non-linearly with $B$. This indicates magnetic anisotropy with  $\Delta S=-1/2$. (b) $\Delta E$ calculated for different values of $\theta$ ($30^{\circ}$ to $90^{\circ}$ every $10^{\circ}$) by numerical diagonalization of the Hamiltonian in equation (\ref{GS}). The parameters are described in the main text. (c) $\text{d}I/\text{d}V$ color plot of the position of the Coulomb peak in $V_{\text{g}}$ as a function of $B$. Each vertical cut is a measurement as shown in (a) in steps of $\Delta B= 50$ mT. (d) $\Delta E=\beta V_{\text{g}}$ versus $B$ measured for different angles of rotation. A vertical offset is substracted to make $\Delta E (B=0)=0$. The angle is changed in-situ using a piezo-driven rotator. Solid lines are fits using equation (\ref{GS}) with parameters $D_{N}=56$ $\mu$eV, $D_{N+1}=69.5$ $\mu$eV, $S_{N}=5$, $S_{N+1}=9/2$. The angles are: Blue: $\theta_N=18\degree$, $\theta_{N+1}=22\degree$. Green: $\theta_N=48\degree$, $\theta_{N+1}=46\degree$. Yellow: $\theta_N=59\degree$, $\theta_{N+1}=56\degree$. Orange: $\theta_N=80\degree$, $\theta_{N+1}=77\degree$.}
 \label{figure6}
 \end{figure}

The detection of magnetic anisotropy through the analysis of SET excitations and inelastic co-tunneling transitions is sensitive to $\Gamma$ and can overlap or coexist with additional low-energy excitations. In contrast, the \emph{gate spectroscopy} method \cite{Burzuri2012} is mostly insensitive to $\Gamma$. The gate spectroscopy focuses on the crossing of the Coulomb edges at zero-bias. This degeneracy point marks the transition from the ground state ($S_{N}$, $m_\text{s}=\pm S_{N}$) of the charge state $N$ to the ground state ($S_{N+1}$, $m'_\textrm{s}=\pm S_{N+1}$) of the adjacent charge state $N+1$ (blue arrow in Fig.\ref{figureSET}(a)). Its position in $V_{\text{g}}$ depends on the energy difference $\Delta E=E_{N+1}-E_{N}$, i.e., the chemical potential $\mu (N+1)$, between those two states. The value of $\Delta E$ changes under an external magnetic field and accordingly the Coulomb peak position shifts in $V_{\textrm{g}}$. This shift is shown in Fig.\ref{figure6}(a) for the Coulomb peak of the Fe$_{4}$ molecular junction in Fig.\ref{figure5} measured at $T=360$ mK. Here, the $\textrm{d}I/\textrm{d}V$ as a function of $V_\textrm{g}$ is measured at zero DC bias voltage with a lock-in modulation of 0.1 mV for different magnetic fields. For isotropic molecules, the shift is linear with $B$ (Zeeman effect):

\begin{equation}
\begin{split}
\Delta E(B)=\beta V_\textrm{g}(B)\\
=E_{N+1}(B)-E_{N}(B) \\
=-g\mu_{\textrm{B}}B\Delta S
\label{GSisotrop}
\end{split}
\end{equation}

\noindent In contrast, for anisotropic molecules, the evolution of $\Delta E$ can be strongly non-linear with $B$ since it also depends on $\theta$. The position of the degeneracy point in $V_\textrm{g}$ shifts accordingly and therefore it can be used to obtain quantitative information of the anisotropy parameters.

Figure \ref{figure6}(b) shows the $B$-dependence of $\Delta E$ calculated with equation (\ref{GS}) for the Fe$_{4}$ SMM  by using different values of $\theta$. In the calculation we assume $D_{N}=D_{N+1}=56~\mu$eV as in the bulk\cite{Accorsi2006}, $\theta_{N}=\theta_{N+1}=\theta$, $S_{N}=5$, $S_{N+1}=9/2$ and $E=0$ for simplicity. For low values of $\theta$ the Zeeman effect is dominant over the anisotropy and therefore $\Delta E$ changes linearly with $B$. For higher values of $\theta$, the magnetic anisotropy overcomes the Zeeman effect at low magnetic fields inducing a flatter $B$-dependence of $\Delta E$. This can be easily deduced by inspecting the magnetic field evolution of the ground states shown in Fig.\ref{figure2}(c,d). The Zeeman dependence is recovered at higher $B$, which for Fe$_{4}$ bulk anisotropy parameters occurs for $B\gtrsim 4$ T. This crossover $B$ between the two regimes is determined by $D$ and can also be used to estimate it. Figure \ref{figure6}(c) shows a color plot of the Coulomb peak position measured as a function of $B$. Each vertical trace is a $\textrm{d}I/\textrm{d}V$ trace as shown in Fig.\ref{figure6}(a) taken at regular steps of $\Delta B= 50$ mT. The center of the peak in $V_\textrm{g}$ is obtained by fitting it with a Lorentzian function. Thereafter $\Delta E=\beta V_{\textrm{g}}$ is calculated with the $\beta$ value determined from the measurements.

Figure \ref{figure6}(d) shows $\Delta E$ measured for different values of $\theta$. A vertical offset is applied to make $\Delta E (B=0)=0$. The angle is changed $in-situ$ by rotating the sample with a piezo-driven positioner. The angular dependence of $\Delta E$ is a direct consequence of the magnetic anisotropy of the Fe$_{4}$ SMM. Solid lines are fits to $\Delta E$ by using equation (\ref{GS}). We assume the most general case in which $D_{N}\neq D_{N+1}$ and $\theta_{N}\neq\theta_{N+1}$. We further assume the left-hand charge state to be the neutral as explained below with $D_{N}=56~\mu$eV as in bulk. From the fits we obtain $D_{N+1}=69.5$ $\mu$eV and the angles shown in Fig.\ref{figure6}(d). The two main conclusions from the fits are: (i) the axial magnetic anisotropy $D$ increases about $24\%$ by adding one electron to the molecule. (ii) The easy axes in adjacent charges states are (almost) collinear. A large misalignment would manifest itself in gate spectroscopy as peaks or dips~\cite{Burzuri2012} around $B=0$ T.

In 9 of the 12 samples, we obtain $D_{N\pm1}$ parameters in the range $60-70~\mu$eV by assuming $D_N=56~\mu$eV. The determination of the neutral state is addressed below. In the other three samples, we obtain smaller ($<60~\mu$eV) $D_{N\pm1}$ parameters. The increment and values of $D$ upon charging the molecule obtained in gate spectroscopy measurements are consistent with the ZFS in the co-tunneling excitations in Sect.\ref{sec:sect5}. Importantly, we note that in sample AB the same value of $\theta$ is obtained independently from the ZFS in the co-tunneling and the gate spectroscopy technique. Moreover, the value of $D\approx70~\mu$eV that yields a ZFS of 0.56 meV is consistent with the ZFS extracted from the co-tunneling excitation. Other samples do not show simultaneously ZFS in the co-tunneling and a clear Coulomb peak and therefore this complementary analysis could not be performed.

In the high-$B$ regime ($B>4$ T), the Zeeman effect is dominant over the anisotropy at any $\theta$ and therefore $\Delta E(B)$ recovers a linear behavior. The sign of the slope provides information about the change in the spin. The slope is positive(negative) for $\Delta S=-1/2$ ($\Delta S=+1/2$) according to equation (\ref{GSisotrop}). The existence of a crossing point at zero bias allows to discard, in this particular case, more complex phenomena like spin blockade ($\Delta S\neq 1/2$) \cite{Heersche2006,Romeike2008}. In the case of Fe$_{4}$, energy considerations and DFT calculations \cite{Nossa2013} predict that the spin ground state of both oxidized and reduced states is $S_{N-1}=S_{N+1}=9/2$. The spin transition will therefore be either $S_{N-1}=9/2$ to $S_{N}=5$, meaning $\Delta S=+1/2$ and negative slope, or $S_{N}=5$ to $S_{N+1}=9/2$, meaning $\Delta S=-1/2$ and positive slope. The sign of the slope can therefore be used to assign the neutral state of the molecule in the stability plots even if no ZFS is detected in the Coulomb blockade regime or the SET. Note that this analysis assumes the theoretical predictions to be correct, i.e., there is no $S=11/2$ charged state. We observe that 8 of the samples show positive slope ($\Delta S=-1/2$) and 4 shows negative slope ($\Delta S=1/2$) at high $B$.


\section{\label{sec:sect7}Gate spectroscopy: Transverse Magnetic Anisotropy.}

\begin{figure}
\includegraphics[width=0.5\textwidth]{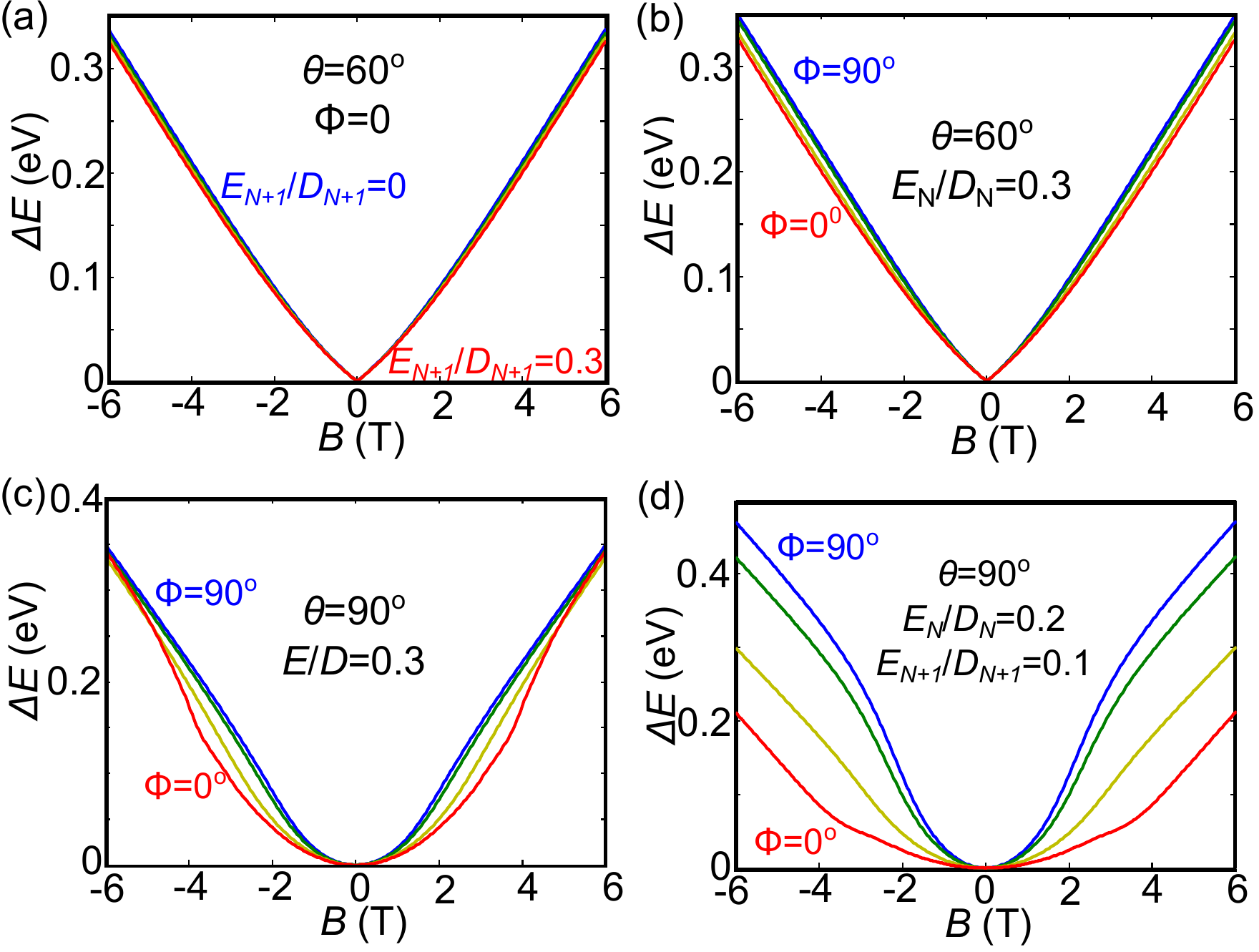}
\caption{$\Delta E$ as a function of the magnetic field calculated by numerical diagonalization of equation (\ref{GS}) by including the transverse anisotropy $E$. For simplicity we take $\phi_{N}=\phi_{N+1}=\phi$ and $\theta_{N}=\theta_{N+1}=\theta$. Other  parameters are: (a) $\theta=60^{\circ}$, $\phi=0$, (b) $\theta=60^{\circ}$, $E_N/D_N=E_{N+1}/D_{N+1}=0.3$; (c) $\theta=90^{\circ}$, $E_N/D_N=E_{N+1}/D_{N+1}=0.3$, (d) $\theta=90^{\circ}$, $E_N/D_N=0.2$ and $E_{N+1}/D_{N+1}=0.1$. The effect of the transverse anisotropy is only relevant when $\theta$ and the ratio $E/D$ are relatively high. Moreover, for $E_N/D_N\neq E_{N+1}/D_{N+1}$ a stronger influence of $E$, thus a stronger dependence on $\phi$ is predicted.}
\label{Transverse}
\end{figure}

So far we have neglected the effect of the transverse magnetic anisotropy in the gate spectroscopy analysis. The magnitude of $E$ is typically more than one order of magnitude lower than $D$ and therefore its effect on the energy spectrum is expected to be subtle. Figure~\ref{Transverse} shows $\Delta E$ calculated by numerical diagonalization of equation (\ref{GS}) for different $\theta$, $E$ and $\phi$ that defines the angle of the magnetic field with the hard anisotropy axis in the plane perpendicular to the easy axis, the so-called hard plane. The values of $D_{N}$, $S_{N}$, $D_{N+1}$ and $S_{N+1}$ are those obtained by gate spectroscopy in Fig.\ref{figure6}. Figures~\ref{Transverse}(a) and (b) show $\Delta E$ calculated for $\theta=60^{\circ}$ and different values of $E$ and $\phi$ respectively. The effect of the transverse anisotropy manifests itself as a slight widening of the $\Delta E$ vs $B$ curve, even for the maximum possible transverse anisotropy $E/D=0.3$. The role of $E$ is therefore negligible even for relatively high values of $\theta$.

In contrast, when the magnetic field is applied close to the hard plane ($\theta\sim 90^{\circ}$), the quantum mixing of the different magnetic states induced by $E$ becomes more relevant and a larger effect in the peak position is predicted as seen in Fig.\ref{Transverse}(c). This effect is maximum when $B$ is applied close to the hard axis ($\phi=0^{\circ}$). Note, however, that only by using high $E/D$ ratios the contribution of $E$ becomes relevant. According to this analysis, for angles $\theta$ far from the hard plane ($\theta\lesssim70^{\circ}$) the gate spectroscopy method introduced in the previous section is more reliable to obtain information on $D$ since the influence of $E$ is small. In contrast, for $\theta\sim90^{\circ}$, and in particular for $\phi\sim 0$, the gate spectroscopy may be used to obtain information of the transverse anisotropy. The large number of free parameters may complicate the analysis in this last case. The scenario becomes even more complicated when we consider $E_{N}/D_{N}\neq E_{N+1}/D_{N+1}$ as shown in Figure~\ref{Transverse}(d). The transverse anisotropy has, in this case, a high impact on the gate spectroscopy that also becomes very sensitive to the angle $\phi$.

\section{\label{sec:sect8}Kondo effect.}

\begin{figure}
\includegraphics[width=0.5\textwidth]{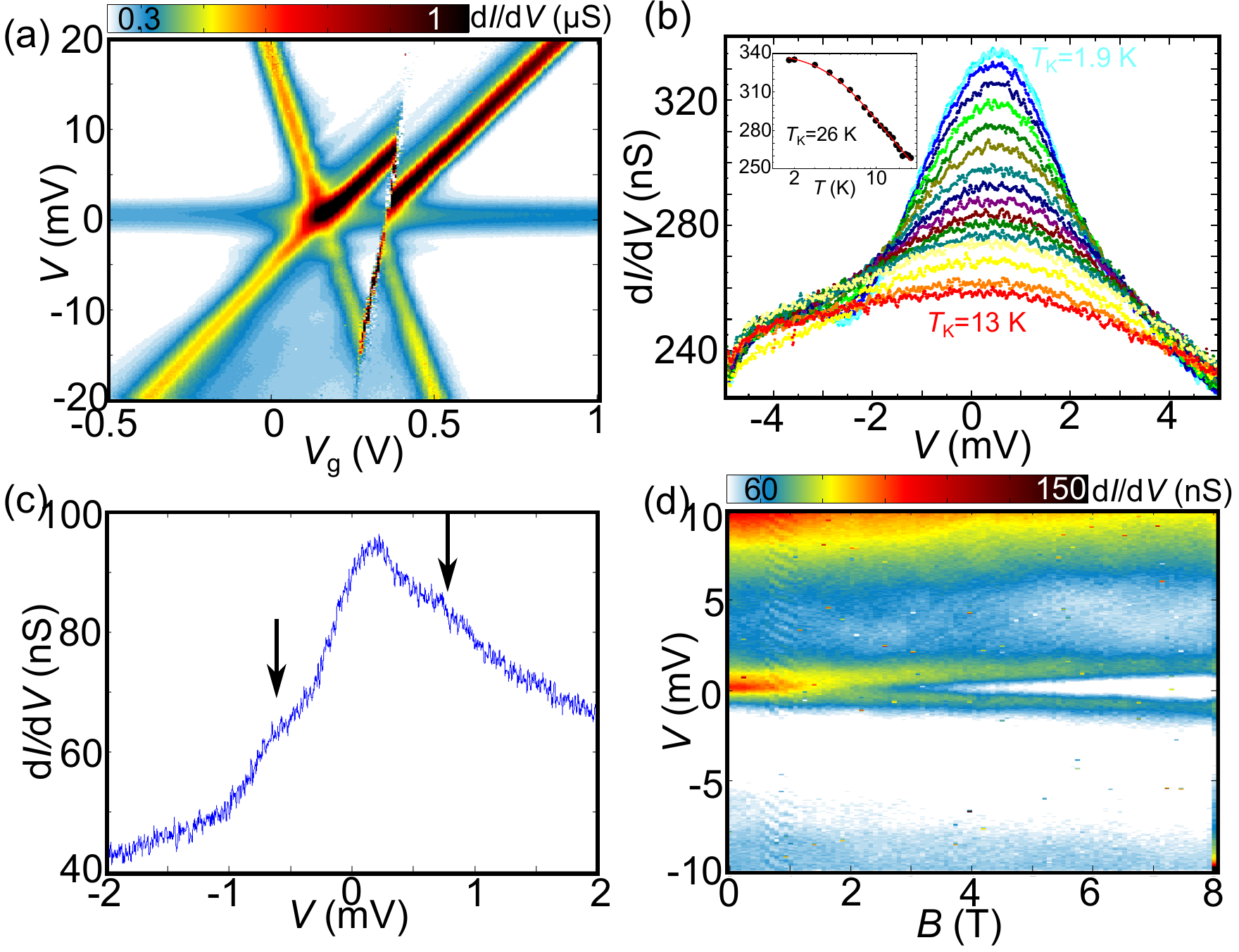}
\caption{(a) $\text{d}I/\text{d}V$ color plot for an Fe$_{4}$ junction measured at $T=1.9$ K showing a zero-bias anomaly in two adjacent charge states (sample F). (b) Temperature dependence of the Kondo resonance. The inset shows the $\text{d}I/\text{d}V$ at $V=0$ as a function of the temperature in logarithmic scale. The low-temperature flattening indicates a strong coupling regime between magnetic core and electrons in which $T_{\text{K}}$ can be accurately determined. The solid line is a fit to an empirical temperature dependence equation with $T_{\text{K}}=26$ K (see text). (c) $\text{d}I/\text{d}V$ trace of a different Fe$_{4}$ junction (sample Y) measured at  $T=370$ mK showing Kondo with $T_{\text{K}}=16$ K plus two additional "shoulders" with energies $\pm0.57$ meV close to the ZFS in bulk. (d) Magnetic field dependence of the Kondo peak shown in (c). The Kondo peak splits following a Zeeman dependence.}
\label{Kondo}
\end{figure}

The Kondo effect is a high-order co-tunneling process in which also the spin of the electron is involved\cite{Kondo1964}. Kondo excitations appear in molecules that have an unpaired electron in one of the orbitals. This unpaired electron can hop into the drain by an elastic process and be replaced by an electron from the source with the opposite spin orientation as illustrated in Fig.\ref{figure3}(d). Kondo physics has been observed in several molecular junctions as a zero-bias peak in the $\textrm{d}I/\textrm{d}V$ in alternating charge states with odd number of electrons\cite{Yu2004,Osorio2007,Scott2010} as sketched in Fig.\ref{figure3}(b). In the case of the Fe$_{4}$, the scenario becomes more complex since there are several unpaired spins in the molecule. Kondo resonances can appear therefore in adjacent charge states which is a fingerprint of the high spin ($S\geq1$) of the molecule\cite{Zyazin2011}. This can be, for example, seen in Fig.\ref{Kondo}(a) in the $\text{d}I/\text{d}V$ color plot for an Fe$_{4}$ junction measured at $T=1.9$ K.

To gain a deeper insight in the nature of this zero-bias excitation we study its evolution in temperature and magnetic field. Kondo correlations appear below a threshold temperature $T_\textrm{K}$ and increase by decreasing the temperature until a regime of strong coupling between spin and electrons is achieved\cite{Zhang2013}. Figure~\ref{Kondo}(b) shows the temperature dependence of the Kondo peak measured at $V_{\textrm{g}}=0.5$ V in the junction of Fig.\ref{Kondo}(a). The height of the peak ($\textrm{d}I/\textrm{d}V_{\textrm{max}}$) decreases with increasing temperature as expected for a Kondo anomaly. In the inset of the figure, $\textrm{d}I/\textrm{d}V_{\textrm{max}}$ is plotted as a function of temperature on a logarithmic scale. The low-temperature flattening of the conductance indicates a strong coupling of the magnetic core with the electrodes ($T<<T_\text{K}$). The experimental curve can be fitted to an empirical $S=1/2$ Kondo temperature dependence:

\begin{equation}
G(T)=G_{\text{b}}+G_{\text{0}}[1+(2^{1/0.22}-1)(T/T_{\text{K}})^{2}]^{-0.22}
\label{EqKondo}
\end{equation}

\noindent with $T_\text{K}=26$ K. $G_\text{0}$ is the conductance in the $T\rightarrow 0$ limit and $G_\text{b}$ is the background contribution. An additional fingerprint of Kondo physics is the splitting of the peak above a threshold magnetic field $B_\text{c}$ followed thereafter by a linear dependence on $B$. For strong coupling between electrons and the magnetic core, this happens when the Zeeman effect overcomes the thermal energy $B_\textrm{c}=(0.5k_\textrm{B}/\mu_\textrm{B})T_\textrm{K}$~\cite{Zhang2013}. In the particular Fe$_{4}$ junction shown in Fig.\ref{Kondo}(b), $B_\textrm{c}=19$ T which is beyond the limitations of the experimental setup and therefore we do not observe any splitting in $B$.

Worth to note is that the intensity of the Kondo peak ($G_\textrm{0}$) is proportional to $\Gamma^2$ and may vary for adjacent charge states\cite{Roch2009}. As explained before, $\Gamma$ varies from sample to sample and therefore the Kondo peak is not always visible in conductance measurements. We have observed Kondo peaks in 12 samples, see Table \ref{TableKondo} for statistics. Excluding sample F described before, the Kondo temperature was determined as the FWHM of the zero-bias anomaly which may give an overestimated $T_\textrm{K}$ due to thermal broadening~\cite{Zhang2013}. Lower $T_\textrm{K}$ values are statistically associated with lower values of $\Gamma$. For instance $\Gamma\textrm{(sample F)}=2$ meV and $\Gamma\textrm{(sample AB)}=0.5$ meV. The presence of a Kondo anomaly may mask the detection of the ZFS as a co-tunneling excitation. In the junction analyzed in Fig.\ref{Kondo}(b), for instance, the FWHM of the Kondo peak is close to 4 meV and therefore the ZFS is not visible. In contrast, in a different Fe$_4$ junction shown in Fig.\ref{Kondo}(c) measured at $T=370$ mK, the Kondo temperature is lower ($T_{\textrm{K}}= 16$ K) and two small "shoulders" can be observed at $V=\pm0.57$ mV on top of the Kondo peak. These values point to the ZFS. Figure \ref{Kondo}(d) shows the magnetic field dependence of the Kondo peak in this junction measured at $T=1.8$ K. The splitting occurs at lower $B$ in this case due to the lower $T_{\textrm{K}}$ and to the fact that the spin/electrons system is not in the strong coupling regime at this temperature\cite{Zhang2013}.

\begin{table}
\begin{tabular}{||c||c||c||}
\hline
\hline
Sample & $T_\textrm{K}$ (K) (left) & $T_\textrm{K}$ (K) (right)\\
\hline
\hline
C & 29 & 29 \\
\hline
D & 23 & 23 \\
\hline
F & 26 & 26\\
\hline
H & - & 30 \\
\hline
I$^+$ & $<45$ & $<45$\\
\hline
N & - & 37 \\
\hline
O & - & 46\\
\hline
Q & - & 20\\
\hline
R & 29 & 29\\
\hline
U &26 & 26\\
\hline
Y & - &16\\
\hline
AB & 4 & 4\\
\hline
\hline
\end{tabular}
\caption{Statistics on Kondo anomalies in Fe$_4$ junctions. The Kondo temperature is obtained from the FWHM of the Kondo peak except for sample F where we use the S=1/2 Kondo temperature dependence. The values for samples with '+' are upper boundaries since additional low-bias excitations hindered the determination of the FWHM. Samples C and D were analyzed in Ref.\cite{Zyazin2011}.}
\label{TableKondo}
\end{table}

The behavior of the Kondo anomaly so far is in good agreement with a $S=1/2$ Kondo model. Interestingly, several theoretical works predict a more exotic Kondo arising from the magnetic anisotropy and high spin of SMM; i.e., high-spin under-screened Kondo and quantum-spin tunneling Kondo \cite{Romeike2006}. Under-screened Kondo has been observed in a C$_{60}$ molecule\cite{Parks2007,Roch2009}, but its experimental observation in SMM with larger spin still remains an open challenge to the best of our knowledge.

\section{\label{sec:sect9}High-energy excitations: Spin excited multiplets.}

\begin{figure}
\includegraphics[width=0.5\textwidth]{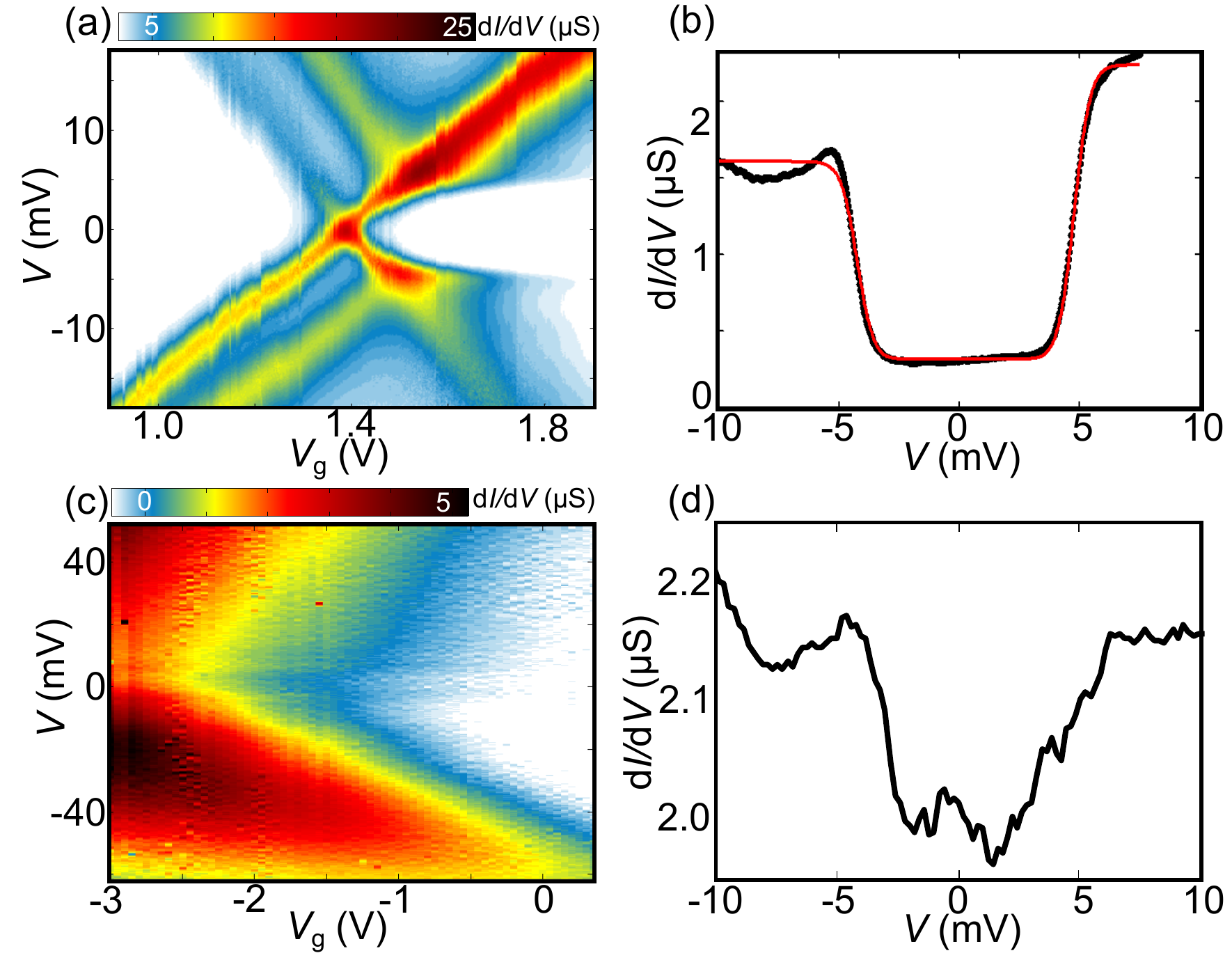}
\caption{(a) $\textrm{d}I/\textrm{d}V$ color map measured versus $V$ and $V_\textrm{g}$ in an Fe$_4$ junction (sample AC) measured at $T=73$ mK. A clear SET excitation shows up at 5 meV that merges at the Coulomb edge with the correspondent co-tunneling excitation in the right-hand charge state. (b) $\textrm{d}I/\textrm{d}V$ trace measured at $V_\textrm{g}=1.9$ V in the Coulomb blockade regime of (a). An inelastic co-tunneling excitation shows up as step in the $\textrm{d}I/\textrm{d}V$ centered at $V=\pm5$ meV. The solid line is a fit to the Lambe-Jaklevic equation. (c) $\textrm{d}I/\textrm{d}V$ color map of a different Fe$_4$ junction (sample P) and (d) $\textrm{d}I/\textrm{d}V$ trace at V$_{\textrm{g}}=-2$ V. An inelastic co-tunneling process shows up centered around $V=\pm5$ meV.}
\label{figure7}
\end{figure}

High-energy excitations, with energies that appear beyond the ground spin multiplet of the Fe$_4$ SMM ($>1.4$ meV), show up in transport (SET and co-tunneling) when triggered by higher bias voltages. In principle, excitations can be either of magnetic or vibrational origin. Magnetic excitations can be identified due to their characteristic magnetic field evolution in contrast with vibrational excitations for which the energy is expected to be magnetic field independent. Figure \ref{figure7}(a) shows the stability plot measured at $T=73$ mK for an Fe$_4$ junction showing a strong excitation at 4.8 meV in the SET that merges in the right diamond edge with the correspondent co-tunneling excitation coming from the right-hand charge state. The energy of the excitation approximately matches the energy in bulk for the $S=4$ excited multiplet of the Fe$_4$ ~\cite{Carretta2004,Mannini2010} and appears, depending on $\Gamma$, as SET or co-tunneling excitations in 8 of the measured samples. Figure \ref{figure7}(b) shows a $\text{d}I/\text{d}V$ trace taken at $V_\text{g}=1.9$ V corresponding to the Coulomb blockade regime in Fig.\ref{figure7}(a). A step in the conductance shows up which is symmetric by reversing the bias polarity. The solid line is a fit to the Lambe-Jaklevic equation (\ref{LJ}) with 4.8 meV. An additional example showing co-tunneling excitations at around 5 meV is shown in Figs.\ref{figure7}(c) and (d) for a different junction containing Fe$_4$. In order to confirm the magnetic origin of these high-bias excitations, further magnetic field measurements are necessary and will by the subject of a further study.

According to the spin selection rules described in Sect.\ref{sec:sect4} and \ref{sec:sect5}, the SET excitation would correspond to a $S_N=5 \rightarrow S_{N+1}=9/2 \rightarrow S_N=4$ transition in the SET regime and a $S_N=5 \rightarrow S_{N}=4$ in the Coulomb blockade. Interestingly, the same selection rules forbid the SET transition involving the excited multiplet of the $S_{N+1}$ charge state, that is  $S_{N+1}=9/2 \rightarrow S_N=5  \rightarrow S_{N+1}=7/2$, since $\Delta S \neq 1/2$. Only one magnetic excitation should be visible in the $\textrm{d}I/\textrm{d}V$ color plot ending in the Coulomb edge that delimits the $S=5$ neutral state whereas it should be absent in the Coulomb edge that delimits the $S=9/2$ charged state. This asymmetry combined with the presence of ZFS and the sign of $\Delta E$ slope in gate spectroscopy, may be used to determine the neutral charge state; i.e., the right-hand one in the specific case shown in Figs.\ref{figure7}(a,c) or the left in Fig.\ref{Kondo}(a). This last example (sample F) is consistent with the positive slope of $\Delta E (B)$ found by gate spectroscopy which also indicates that the left charge state is the neutral one\cite{footnote2}.

\section{\label{sec:sect10}Vibrational modes in individual magnetic molecules. Strong coupling.}

\begin{figure}
\includegraphics[width=0.5\textwidth]{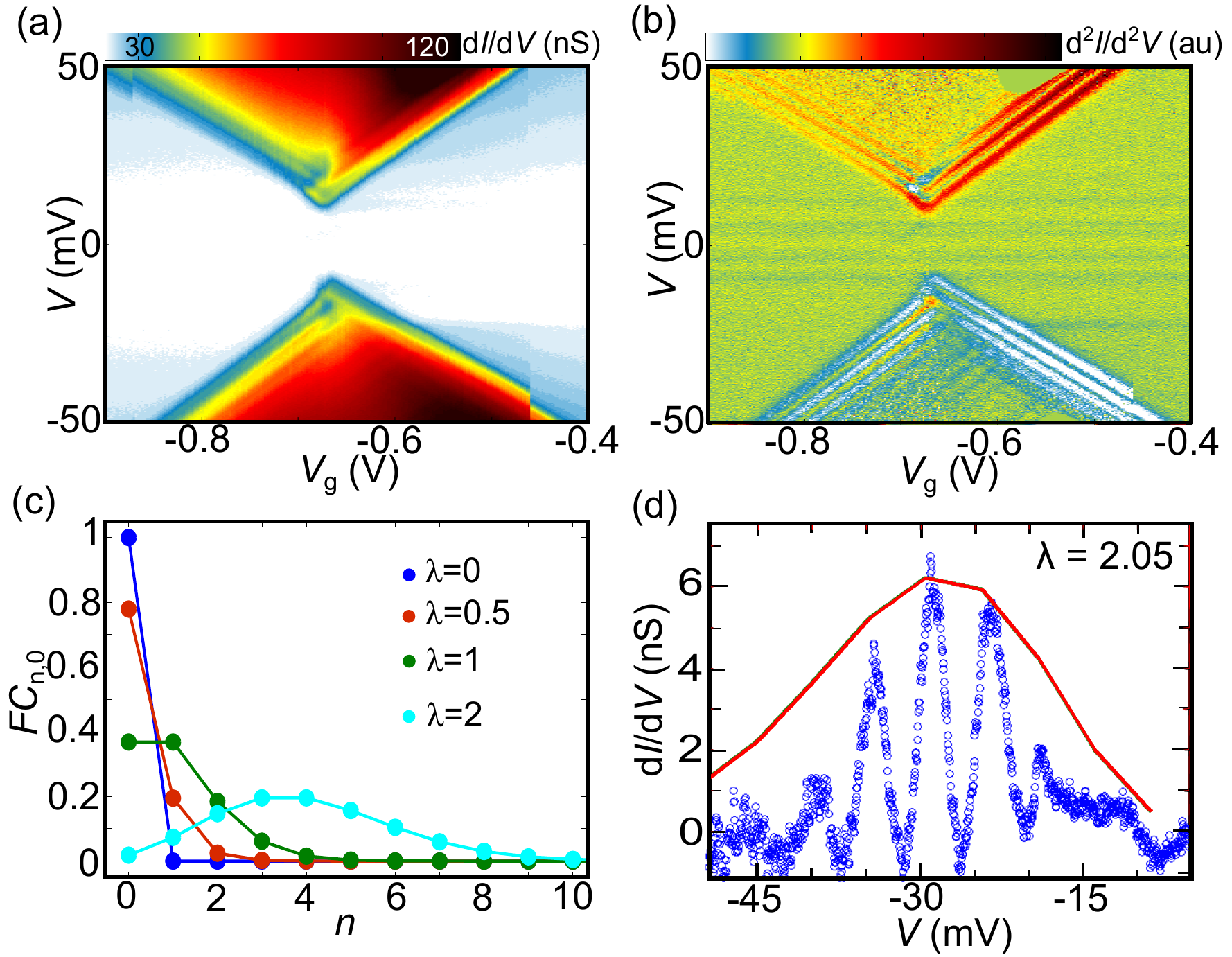}
\caption{(a) $\textrm{d}I/\textrm{d}V$ color plot as a function of $V$ and $V_{\textrm{g}}$ for an Fe$_{4}$ junction (sample L) measured at $T=1.8$ K. Several equally-spaced excitations and a low-bias suppression of the SET are observed. These are fingerprints of Franck-Condon blockade due to a high electron-phonon coupling. This figure is adapted from Ref.\cite{Burzuri2014}.  (b) Numerical derivative of $\textrm{d}I/\textrm{d}V$ color plot shown in (a). Vibrational excitations become more visible. (c) Franck-Condon factors for different electron-phonon couplings $\lambda$. Higher harmonics $n$ are excited for large $\lambda$ values. (d) $\text{d}I/\text{d}V$ trace at a fixed $V_{\textrm{g}}$. This figure is adapted from Ref.\cite{Burzuri2014}}.
\label{FigureFC}
\end{figure}

Vibrations are present in all kind of nanostructures and can play an important role in their mechanics. In addition, vibrational modes can couple to the charge carriers and therefore show up in conductance measurements. Figure~\ref{FigureFC}(a) shows the $\textrm{d}I/\textrm{d}V$ color plot of an Fe$_{4}$ junction in which vibrations appear as SET lines parallel to the diamond edges\cite{Burzuri2014}. The measurements are performed at $T=1.8$ K. The excitations can also be observed in the conductance plot in Figs.\ref{figure5}(a) and \ref{figure5}(b). The energies of vibrational excitations are independent of the magnetic field in contrast with magnetic excitations. Moreover, the energy is also independent of the bias polarity and the charge state. Figure~\ref{FigureFC}(b) shows the numerical derivative of the $\textrm{d}I/\textrm{d}V$ color plot in Fig.\ref{FigureFC}(a) in which vibrational excitations become more visible.

The strength of the electron-phonon coupling $\lambda$ determines the $\text{d}I/\text{d}V$ height of the excitations ($\text{d}I/\text{d}V\mid_\text{max}$). In short, $\text{d}I/\text{d}V\mid_\text{max}$ is proportional to the Franck-Condon factors $FC_{n,0}=\lambda^{2n}e^{-\lambda^2}/n!$ where $n$ is the quantum state of the vibration in an harmonic oscillator approximation. Figure~\ref{FigureFC}(c) shows the Franck-Condon factors for different values of $\lambda$ for transitions between the ground vibrational state and higher harmonics. Intermediate and high values of $\lambda$ excite higher harmonics of the vibration that appear in the conductance when the bias matches a quantum of the energy of the vibration $n\hbar\omega$. This is the case in Fig.\ref{FigureFC}(b) where at least $n=8$ harmonics can be observed. The regular energy spacing of the excitations indicates that a single vibrational mode is the origin of the spectrum. The energy spacing, taken from the intersection of the excitations with the diamond edge, is the energy of the vibrational mode and equals $\hbar\omega=2.6$ meV~\cite{Burzuri2014}. Even if other vibrational modes of the molecules are excited, only those with the higher $\lambda$ will contribute significantly to the current~\cite{Burzuri2014}.

For higher values of $\lambda$, the FC factor for ground state to ground state transitions ($FC_{0,0}$) decreases exponentially and becomes negligible as seen in Fig.\ref{FigureFC}(c). Low-bias SET transitions are therefore suppressed and a low-bias gap appears in the conductance that cannot be lifted with a gate voltage, as shown in Fig.\ref{FigureFC}(a). Only by increasing $V$, higher harmonics become available enabling transport and lifting the blockade. This vibrational-induced suppression of the current is known as Franck-Condon blockade.

The value of $\lambda$ can be estimated from the relative height of the different harmonics of the oscillation. Figure~\ref{FigureFC}(d) shows a $\text{d}I/\text{d}V$ trace measured at a fixed $V_\text{g}=-0.711$ V. The background contribution due to direct tunneling in the junction is subtracted from the data. The solid line is a fit to the Franck-Condon factor $FC_{n,0}$ with $\lambda=2.05$. An independent estimate of $\lambda$ can be obtained from the size of the gap as $V_\text{th}=\lambda^{2}\hbar\omega$ where $V_\text{th}$ is the threshold bias voltage that restores SET. The value obtained from Fig.\ref{FigureFC}(a) is $\lambda=1.7$.

The value of $\lambda$ may vary from sample to sample due to environmental factors like image charges and different molecular orientations relative to the electrodes \cite{Pasupathy2005,Osorio2007,Burzuri2014}. This may explain why the Franck-Condon blockade is not present in all junctions. We have observed vibrational excitations in 10 samples and a low-bias suppression of the current in three of them. Vibrational energies are in the order of 2 meV. We further note that our measurements show that $\lambda$ can be high in SMM. It is unclear how such a strong electron-phonon coupling affects the magnetic properties of the Fe$_{4}$. For instance, vibrations are known to be a source of spin decoherence\cite{Stamp2009} and may lead to quantum interferences\cite{Zhong2009}.

\section{\label{sec:sect11}Conclusions}

We have reviewed different methods to detect and measure magnetic properties of single molecules in electronic transport. We have focussed on the Fe$_4$ SMM and provided statistics of the relevant magnetic properties of the molecule. We find that the coupling of the molecules to the electrodes $\Gamma$, the gate coupling parameter $\beta$ and the electron-phonon coupling vary from sample to sample which may be explained by a change in the arrangement of the molecule between the electrodes. This variation induces changes in the perceptibility of the magnetic and vibrational excitations or Kondo anomalies. However, the magnetic properties of the molecule like the high-spin and the axial magnetic anisotropy are preserved, including a reversible increase in the magnetic anisotropy upon charging of the molecule. These results point to the Fe$_4$ SMM as a good platform to study the interesting physics arising from the spin-charge interaction at the single-molecule level. Quantum effects, such as quantum tunnel or quantum oscillations induced by the transverse anisotropy may have an impact on the current through the molecule. In addition, further studies are needed to shed more light into the spin-vibration interaction or the existence of exotic Kondo behavior due to the high spin of the molecule.

\begin{acknowledgments}
This work was supported by the EU FP7 program through project 618082 ACMOL and an ERC grant advance (Mols@Mols) and by the Dutch funding organizations OCW and NWO(VENI).
\end{acknowledgments}


\begin{thebibliography}{33}
\bibitem{footnote1} We label the samples chronologically with a capital letter or two capital letters after sample Z. We also indicate the samples that have been reported in previous publications.
\bibitem{footnote2} Unfortunately, no gate spectroscopy was carried out for samples AC and P.
\end{thebibliography}
\end{document}